\pdfoutput=1

\documentclass[11pt,twoside,a4paper,cmspaper,final,collab]{cms-tdr}
\def\svnVersion{488084P}\def\cmsCernNoTag{CERN-EP-2018-230}\def\cmsCernDate{\today}\def\cmsMessage{Published in the European Physical Journal C as \href{http://dx.doi.org/10.1140/epjc/s10052-019-6556-3}{\doi{10.1140/epjc/s10052-019-6556-3.}}}

\begin{document}\cmsNoteHeader{B2G-17-018}

\hyphenation{had-ron-i-za-tion}
\hyphenation{cal-or-i-me-ter}
\hyphenation{de-vices}
\RCS$HeadURL: svn+ssh://svn.cern.ch/reps/tdr2/papers/B2G-17-018/trunk/B2G-17-018.tex $
\RCS$Id: B2G-17-018.tex 482472 2018-11-27 10:30:49Z rkogler $

\providecommand{\cmsTable}[1]{\resizebox{\textwidth}{!}{#1}}
\newcommand{\HTlep}{\ensuremath{S_{\mathrm{T},\text{lep}}}\xspace}
\newcommand{\pb}{\unit{pb}}
\newcommand{\msd}{\ensuremath{m_{\mathrm{SD}}}\xspace}
\newcommand{\mreco}{\ensuremath{m_\mathrm{reco}}\xspace}
\newcommand{\mVLQ}{\ensuremath{m_\mathrm{{VLQ}}}\xspace}
\newcommand{\mtop}{\ensuremath{m_{\cPqt}}\xspace}
\newcommand{\mW}{\ensuremath{m_{\PW}}\xspace}
\newcommand{\mtopmean}{\ensuremath{\overline{m}_{\cPqt}}\xspace}
\newcommand{\mWmean}{\ensuremath{\overline{m}_{\PW}}\xspace}
\newcommand{\ptW}{\ensuremath{p_{\mathrm{T},\PW}}\xspace}
\newcommand{\ptt}{\ensuremath{p_{\mathrm{T},\cPqt}}\xspace}
\newcommand{\akf}{AK4 jet}
\newcommand{\ake}{AK8 jet}
\newcommand{\X}{\ensuremath{X_{5/3}}}
\providecommand{\NA}{\ensuremath{\text{---}}\xspace}
\providecommand{\CL}{CL\xspace}
\newlength\cmsFeyWidth
\ifthenelse{\boolean{cms@external}}{\setlength\cmsFeyWidth{0.35\columnwidth}}{\setlength\cmsFeyWidth{0.35\textwidth}}
\newlength\cmsFigWidth
\ifthenelse{\boolean{cms@external}}{\setlength\cmsFigWidth{0.9\columnwidth}}{\setlength\cmsFigWidth{0.45\textwidth}}
\ifthenelse{\boolean{cms@external}}{\providecommand{\cmsLeft}{top}}{\providecommand{\cmsLeft}{left}}
\ifthenelse{\boolean{cms@external}}{\providecommand{\cmsRight}{bottom}}{\providecommand{\cmsRight}{right}}

\cmsNoteHeader{B2G-17-018}

\title{Search for single production of vector-like quarks decaying to a top quark and a \PW\ boson in proton-proton collisions at $\sqrt{s} = 13 \TeV$}
\titlerunning{Search for vector-like quarks at $\sqrt{s}=13\TeV$}

\date{\today}

\abstract{
A search is presented for the single production of vector-like quarks in proton-proton
collisions at $\sqrt{s}=13\TeV$. The data, corresponding to an integrated luminosity of 35.9\fbinv, 
were recorded with the CMS experiment at the LHC.
The analysis focuses on the vector-like quark decay into a top quark and a $\PW$ boson, with
one muon or electron in the final state. The mass of the vector-like quark candidate
is reconstructed from hadronic jets, the lepton, and the missing transverse momentum.
Methods for the identification of $\cPqb$ quarks and of highly Lorentz boosted hadronically decaying top quarks and
$\PW$ bosons are exploited in this search.
No significant deviation from the standard model background expectation is observed.
Exclusion limits at 95\% confidence level are set on the product of the production
cross section and branching fraction as a function of the vector-like quark mass,
which range from 0.3 to 0.03\pb for vector-like quark masses of 700 to 2000\GeV.
Mass exclusion limits up to 1660\GeV are obtained, depending on the vector-like quark type,
coupling, and decay width. These represent the most stringent exclusion limits for the single production of vector-like quarks in this channel.}

\hypersetup{
pdfauthor={CMS Collaboration},
pdftitle={Search for single production of vector-like quarks decaying to a top quark and a W boson in proton-proton collisions at sqrt(s) = 13 TeV},
pdfsubject={CMS},
pdfkeywords={CMS, physics, vector-like quarks}}

\maketitle

\section{Introduction}
\label{sec:introduction}

The discovery of the Higgs boson (\PH)~\cite{Aad:2012tfa,Chatrchyan:2012xdj}
with a mass of 125\GeV completes the particle content of the standard model (SM).
Even though the SM yields numerous accurate predictions, there are several open questions,
among them the origin of the $\PH$ mass stability at the
electroweak scale. Various models beyond the SM have been proposed that
stabilise the $\PH$ mass at the measured value;
some examples are Little Higgs~\cite{ArkaniHamed:2002qy,Schmaltz:2002wx,Schmaltz:2005ky} or
Composite Higgs models~\cite{Marzocca:2012zn}, in which additional top quark partners
with masses at the TeV scale are predicted. Since the left- (LH) and right-handed (RH) chiral components
of these particles transform in the same way under the SM electroweak
symmetry group, they are often referred to as ``vector-like quarks'' (VLQs).
In contrast to a fourth chiral quark generation, their impact on the
$\PH$ properties is small, such that VLQs have not been excluded by the measurements of $\PH$
mediated cross sections~\cite{Djouadi:2012ae,VLQHandbook:2013,Khachatryan:2016vau}.

Several searches for VLQs have been performed at the CERN LHC, setting lower exclusion limits on the VLQ
mass \mVLQ~\cite{Chatrchyan:2013uxa,Khachatryan:2015axa,Khachatryan:2015oba,Khachatryan:2016vph,Sirunyan:2017tfc,
Sirunyan:2017jin,Sirunyan:2017usq,Sirunyan:2017pks,Sirunyan:2017ynj,Sirunyan:2018fjh,
Aad:2011yn,Aad:2012bdq,Aad:2014efa,Aad:2015gdg,Aad:2015kqa,Aad:2015mba,Aaboud:2016lwz,Aad:2016qpo,Aad:2016shx,Aaboud:2017qpr,Aaboud:2017zfn,
Aaboud:2018xuw}.
Many of these analyses study the pair production of VLQs via the strong interaction.
In contrast, the analysis presented here searches for the single VLQ production via the weak
interaction, where a hadronic jet is emitted at a low angle with respect to the beam direction.
Furthermore, VLQs with enhanced couplings to the third generation quarks (\ie VLQ \PQB and \X\ quarks with an electric charge of $1/3$ and $5/3$ respectively) are produced in association with a bottom ($\cPqb$) or top ($\cPqt$) quark, leading to the \PQB{}+\cPqb, \PQB{}+\cPqt, and \X+\cPqt\ production modes.

\begin{figure}[bh]
  \centering
  \includegraphics[width=\cmsFeyWidth]{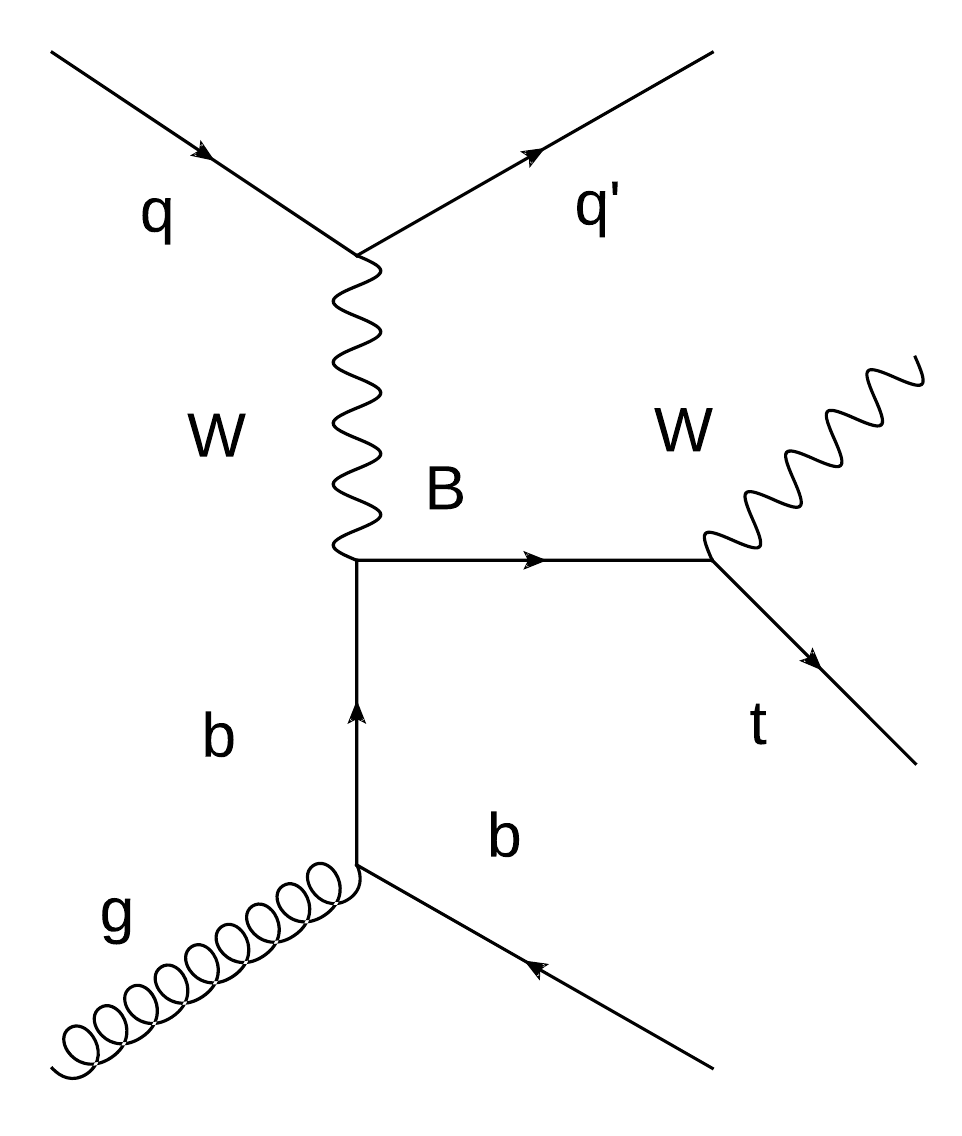}\hspace*{0.08\columnwidth}
  \includegraphics[width=\cmsFeyWidth]{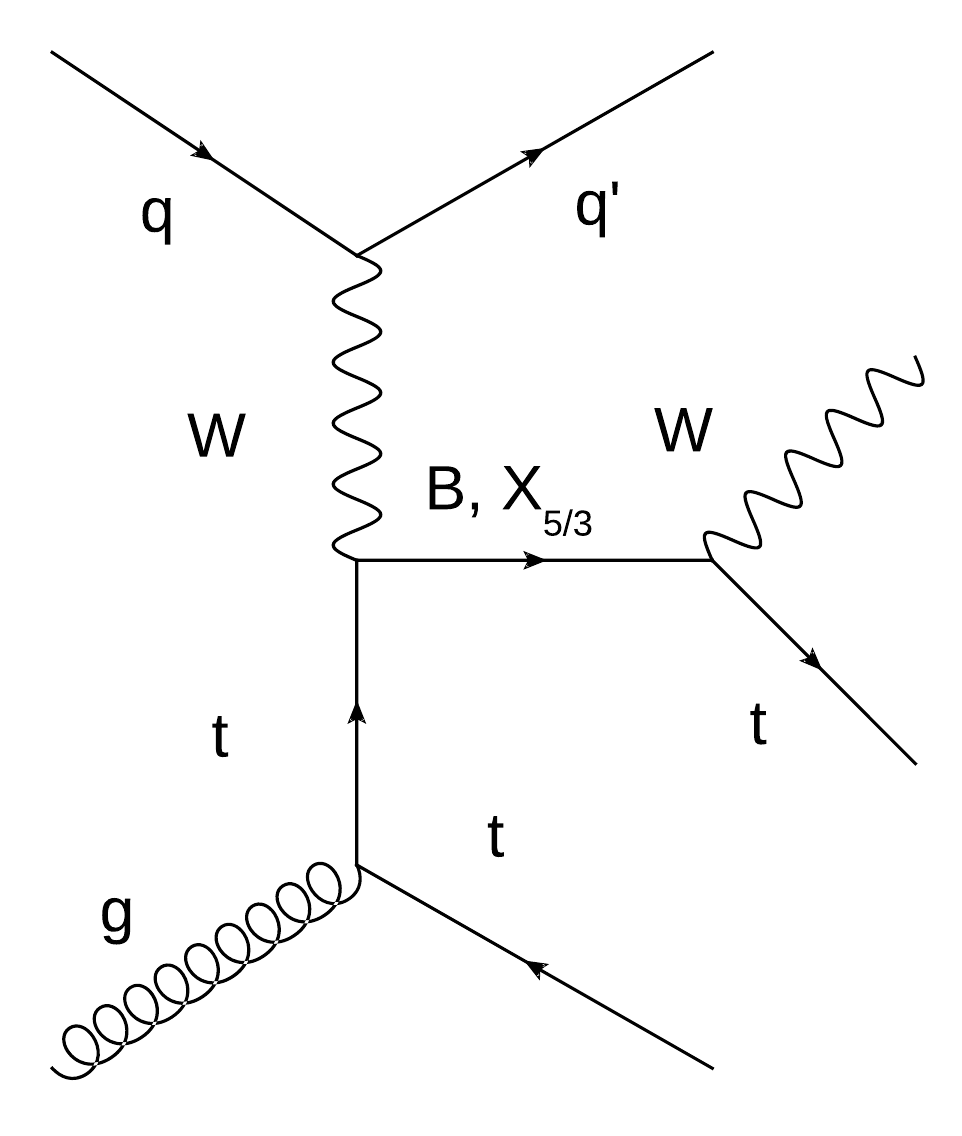}
  \caption{Leading order Feynman diagrams for the production
  of a single vector-like \PQB or $\X$ quark in association
  with a $\cPqb$ (left) or $\cPqt$ (right) and a light-flavour quark,
  and the subsequent decay of the VLQ to $\cPqt\PW$.}
  \label{fig:feynman}
\end{figure}

While a VLQ \PQB quark could decay into the $\PH\cPqb$, $\PZ\cPqb$, or
$\cPqt\PW$ final state, a VLQ \X\ quark could only decay into the \cPqt\PW\ final state. This search focuses on the \cPqt\PW\ final state.
In Fig.~\ref{fig:feynman}, two leading-order (LO) Feynman diagrams are shown for the single production of \PQB and \X\ quarks and their decay into \cPqt\PW. This paper presents the first search of this signature in proton-proton ($\Pp\Pp$) collision data recorded at a centre-of-mass energy of 13 \TeV. Results at $\sqrt{s} = 8 \TeV$ have been obtained by the ATLAS collaboration~\cite{Aad:2015voa}.

In this analysis, final states with a single muon or electron, several hadronic jets, and missing transverse momenta $\ptmiss$ are studied.
Because of the high mass of the VLQ, the $\cPqt$ and $\PW$ can have high Lorentz boosts,
leading to highly collimated decays of the $\PW$ boson, the top quark and non-isolated leptons.
For signal events, the mass of the \PQB and $\X$ quarks can be reconstructed
using hadronic jets, the lepton, and the $\ptmiss$. The associated \cPqb\ and \cPqt, as well as the leptons originating from their decay,
have much lower transverse momenta $\pt$ and are not considered for the reconstruction or selection.

The dominant SM background processes are top quark pair (\ttbar) production,
$\PW$+jets and $\PZ$+jets production, single $\cPqt$ production, and multijet production via the strong force.
All SM backgrounds contributing to this search are predicted from dedicated control regions
in data, defined through the absence of a forward jet.

This paper is organised as follows: Section~\ref{sec:cmsdec} provides a description of the CMS detector.
Section~\ref{sec:samples} introduces the data set and the simulated events.
This is followed by the event selection in Section~\ref{sec:object}, as well as by
the description of the reconstruction of the VLQ mass in Section~\ref{sec:reco}.
In Section~\ref{sec:background}, a method to estimate the background is discussed.
Systematic uncertainties are detailed in Section~\ref{sec:systematic}.
The final results of the analysis, as well as the statistical
interpretation in terms of exclusion limits, are discussed in Section~\ref{sec:results}.

\section{The CMS detector and physics objects}
\label{sec:cmsdec}

The central feature of the CMS apparatus is a superconducting solenoid of 6\unit{m} internal diameter, providing a magnetic field of 3.8\unit{T}. Within the solenoid volume are a silicon pixel and strip tracker, a lead tungstate crystal electromagnetic calorimeter (ECAL), and a brass and scintillator hadron calorimeter (HCAL), each composed of a barrel and two endcap sections. Forward calorimeters extend the pseudorapidity coverage provided by the barrel and endcap detectors. Muons are detected in gas-ionisation chambers embedded in the steel flux-return yoke outside the solenoid.

{\tolerance=800
The particle-flow event algorithm~\cite{Sirunyan:2017ulk} aims to reconstruct and identify each individual particle with an optimised combination of information from the various elements of the CMS detector. The energy of photons is directly obtained from the ECAL measurement, corrected for zero-suppression effects. The energy of electrons is determined from a combination of the electron momentum at the primary interaction vertex, the energy of the corresponding ECAL cluster, and the energy sum of all bremsstrahlung photons spatially compatible with originating from the electron track~\cite{Khachatryan:2015hwa}. The energy of muons is obtained from the curvature of the corresponding track~\cite{Chatrchyan:2012xi}. The energy of charged hadrons is determined from a combination of their momentum measured in the tracker and the matching ECAL and HCAL energy deposits, corrected for zero-suppression effects and for the response function of the calorimeters to hadronic showers. Finally, the energy of neutral hadrons is obtained from the corresponding corrected ECAL and HCAL energy.
\par}

{\tolerance=800
The reconstructed vertex with the largest value of summed physics-object $\pt^2$ is taken to be the primary $\Pp\Pp$ interaction vertex. The physics objects used are the jets, clustered with the jet finding algorithm~\cite{Cacciari:2008gp,Cacciari:2011ma} with the tracks assigned to the vertex as inputs, and the associated missing transverse momentum, taken as the negative vector sum of the \pt of those jets.
\par}

A more detailed description of the CMS detector, together with a definition of the coordinate system used and the relevant kinematic variables, can be found in Ref.~\cite{Chatrchyan:2008zzk}.

\section{Data and simulated samples}
\label{sec:samples}
In this analysis, $\Pp\Pp$ collision data at a centre-of-mass energy of
$13\TeV$ taken in 2016 by the CMS experiment are analyzed.
The data have been collected with muon and electron triggers~\cite{Khachatryan:2016bia}.
For the muon trigger, a muon candidate with $\pt > 50 \GeV$ is required.
Data events in the electron channel are collected using a logical combination
of two triggers: the first requires an electron candidate
with $\pt > 45\GeV$ and a hadronic jet candidate with
$\pt > 165\GeV$, the second requires an electron candidate with $\pt > 115\GeV$.
In the trigger selection, reconstructed leptons and
jets must be in the central part of the detector, with a pseudorapidity of $\abs{\eta} < 2.4$.
No lepton isolation criteria are applied at the trigger level.
The collected data correspond to an integrated luminosity of 35.9\fbinv~\cite{CMS-PAS-LUM-17-001}.

For the study of dominant SM background processes and for the validation of the background estimation,
simulated samples using Monte Carlo (MC) techniques are used.
The top quark pair production via the strong interaction and single top quark production
in the $t$-channel, and the $\cPqt\PW$ process are generated with the next-to-leading-order (NLO)
generator \POWHEG~\cite{Nason:2004rx,Frixione:2007vw,Alioli:2010xd}
(version v2 is used for the first two and version v1 for the third).
The event generator \MGvATNLO(v2.2.2)~\cite{Alwall:2014hca} at NLO is used for single top quark
production in the $s$-channel. The $\PW$+jets and $\PZ$+jets processes are also simulated
using \MGvATNLO(v2.2.2). The $\PW$+jets events are generated at
NLO, and the FXFX scheme~\cite{Frederix:2012ps} is used to match the parton shower emission.
The $\PZ$+jets events are produced at LO with the MLM parton matching scheme~\cite{Alwall:2007fs}.
The production of quantum chromodynamics  (QCD) multijet events has been
simulated at LO using \PYTHIA~\cite{Sjostrand:2014zea}.
All generated events are interfaced with \PYTHIA for the description of the parton
shower and hadronisation.
The parton distribution functions (PDFs) are taken from the NNPDF 3.0~\cite{Ball:2014uwa}
sets, with the precision matching that of the matrix element calculations.
The underlying event tune is CUETP8M1~\cite{Khachatryan:2015pea,Skands2014},
except for the simulation of top quark pairs and single top quark production in the $t$-channel,
which use CUETP8M2T4~\cite{CMS-PAS-TOP-16-021}.

{\tolerance=3000
Signal events are generated at LO using \MGvATNLO for \PQB and $\X$ with VLQ decay widths
relative to the VLQ mass of $(\Gamma/m)_{\mathrm{VLQ}} = 1$, 10, 20, and 30\%.
The samples with 1\% relative VLQ width are simulated in steps of 100\GeV for masses between 700 and 2000\GeV.
Samples with 10, 20, and 30\% relative VLQ widths are generated in steps of 200\GeV for masses
ranging from 800 to 2000\GeV, using a modified version of the
model proposed in Refs.~\cite{Buchkremer:2013bha,Fuks:2016ftf,Oliveira:2014kla}.
Separate signal samples are generated for the two main production modes, in which VLQs 
are produced in association either with a \cPqb\ quark or with a \cPqt\ quark, 
viz.\ $\Pp\Pp \to \PQB \cPqb \cPq$ and $\Pp\Pp \to \PQB \cPqt \cPq$.
The theoretical cross sections for VLQ production are calculated
using Refs.~\cite{Campbell:2004ch,Matsedonskyi:2014mna,Carvalho:2018jkq},
where a simplified approach is used to provide a model-independent
interpretation of experimental results for narrow and large mass width scenarios, as already used for the
interpretation of singly produced vector-like \PQT and \PQB quarks~\cite{Sirunyan:2017ynj,Sirunyan:2018fjh}.
The $\textsc{MADSPIN}$ package~\cite{Frixione:2007zp,Artoisenet:2012st}
is used to retain the correct spin correlations of the top quark and $\PW$ boson decay products. 
Interference effects between signal and SM processes have been found to be negligible in this analysis.
\par}

All generated events are passed through a \GEANTfour~\cite{AGOSTINELLI2003250}
based detector simulation of the CMS detector. Additional $\Pp\Pp$ interactions originating
from the same bunch crossing (in-time pileup), as well as from the following or
previous bunch crossings (out-of-time pileup) are taken into account in the simulation.

\section{Event selection}
\label{sec:object}
The physics objects used in this analysis are muons, electrons, hadronic jets, $\ptvecmiss$,
and \HTlep (defined as the scalar sum of the lepton $\pt$ and $\ptmiss$).

For each event, jets are clustered from reconstructed particles using the
infrared and collinear safe anti-\kt algorithm~\cite{Cacciari:2008gp} with a
distance parameter $R=0.4$ (\akf). Additionally, jets with $R=0.8$ (\ake) are also clustered in
every event with the anti-\kt algorithm, which are used for $\cPqt$ and $\PW$ tagging.
The jet clustering is performed with the \FASTJET~\cite{Cacciari:2011ma} package.
Jet momentum is determined as the vectorial sum of all particle momenta in the jet,
and is found from simulation to be within 5--10\% of the true momentum over the whole \pt
spectrum and detector acceptance. Additional $\Pp\Pp$ interactions within the same or
nearby bunch crossings can contribute additional tracks and calorimetric energy depositions
to the jet momentum. To mitigate this effect, tracks identified to be originating from
pileup vertices are discarded, and an offset correction is applied to correct for
remaining contributions. Jet energy corrections are derived from simulation studies
so that the average measured response of jets becomes identical to that of particle level jets. In situ
measurements of the momentum balance in dijet, photon+jet, $\PZ$+jet, and multijet events
are used to account for any residual differences in the jet energy scale in data and simulation.
Additional selection criteria are applied to each jet to remove jets potentially dominated
by anomalous contributions from various subdetector components or reconstruction
failures~\cite{Khachatryan:2016kdb}.

From the corrected and reconstructed \akf$\mathrm{s}$, those are considered that have $\pt > 30 \GeV$ and $\abs{\eta} < 4$,
while \ake$\mathrm{s}$ must have $\pt>170 \GeV$ and $\abs{\eta} < 2.4$.

Events selected in the analysis are required to have one reconstructed
muon or electron with $\pt >55 \GeV$ and
$\abs{\eta} < 2.4$. Electrons and muons are selected using tight quality
criteria with small misidentification probabilities of about 0.1\% for muons
and 1\% for electrons~\cite{Khachatryan:2015hwa, Sirunyan:2018fpa}.
In the electron channel, a \akf\ must have
$\pt>185 \GeV$ and $\abs{\eta} < 2.4$ if the electron has $\pt <120 \GeV$,
reflecting the trigger selection.
Events with more than one muon or electron passing the same tight identification
criteria and having $\pt > 40 \GeV$ and $\abs{\eta} < 2.4$ are discarded.
Selected events contain two \akf$\mathrm{s}$ with $\pt > 50 \GeV$, which are in the central part of the detector with $\abs{\eta} < 2.4$.
Additionally at least one \ake\ is required. For the reconstruction \akf$\mathrm{s}$ are used with $\pt > 30 \GeV$ and $\abs{\eta} < 2.4$,
while the \akf$\mathrm{s}$ emitted close to the beam pipe and employed in the background estimation must fulfill $\pt > 30 \GeV$ and $ 2.4 < \abs{\eta} < 4$.

Because of the high Lorentz boosts of the top quarks and $\PW$ bosons from the heavy VLQ decay,
signal events can have leptons in close vicinity to the jets.
For this reason, standard lepton isolation would reduce the selection
efficiency considerably. Therefore, for the suppression of events originating
from QCD mulitjet processes, either the perpendicular component of the lepton momentum relative to the geometrically closest \akf\ $p_{\mathrm{T,rel}}$, is required to
exceed $40 \GeV$ or the angular distance of the lepton to the jet,
$\Delta R (\ell,\mathrm{jet}) = \sqrt{\smash[b]{(\Delta \eta)^2 + (\Delta \phi)^2}}$, must be larger than 0.4,
where $\phi$ is the azimuthal angle in radians. Furthermore, for selecting an event, the magnitude of
$\ptvecmiss$ has to be greater than 50\GeV in the muon channel and greater than 60\GeV in the electron channel.
This requirement reduces the amount of background from multijet production.
The final selection is based on the variable \HTlep,
which is required to be larger than 250\GeV in the muon channel and 290\GeV in the electron channel.

Events are separated into categories exploiting the tagging techniques for
boosted top quarks and $\PW$ bosons decaying hadronically, as well as for hadronic jets originating from $\cPqb$ quarks.
Jets with $R=0.8$ are used to identify the hadronic decays of highly boosted top quarks and
$\PW$ bosons~\cite{CMS:2016tvk,CMS-PAS-JME-16-003}. 
For top quark jets $\pt>400 \GeV$ is required,
and for $\PW$ boson jets the requirement is $\pt>200 \GeV$.
The ``soft drop'' (SD) declustering and grooming algorithm~\cite{Dasgupta:2013ihk, Larkoski:2014wba}
with $z=0.1$ and $\beta=0$ is employed to identify subjets and to remove soft and
wide-angle radiation.
The groomed jet mass, \msd, is used to identify top quark and $\PW$ boson candidates.
Tagged top quark candidates ($\cPqt$ tagged) are required to have $105 < \msd < 220 \GeV$
and one of the subjets must fulfill the loose $\cPqb$ tagging criterion, based on
the combined secondary vertex (CSVv2)~\cite{Sirunyan:2017ezt} algorithm. The loose
criterion is defined to give a 80\% efficiency
of correctly identifying \cPqb\ jets, with a 10\% probability of incorrectly tagging a light quark jet.
Additionally, the jet must have a
N-subjettiness~\cite{Thaler:2010tr,Thaler:2011gf} ratio $\tau_3/\tau_2 < 0.5$ and
its angular distance to the lepton $\Delta R{(\ell,\cPqt\ \mathrm{tag})}$
must be larger than 2.
Identified $\PW$ boson candidates ($\PW$ tag) must have $65 < \msd < 95 \GeV$.
The medium $\cPqb$ tag criterion is used on \akf$\mathrm{s}$, defined to give a $60\%$ efficiency of correctly identifying \cPqb\ jets,
with a $1\%$ probability of incorrectly tagging a light quark jet.

Selected events are attributed to different mutually exclusive event categories.
Events containing at least one $\cPqt$ tag constitute the first category
(``\cPqt\ tag''). If no $\cPqt$ tag is found,
all events with at least one $\PW$ tag
are grouped into a second category (``\PW\ tag'').
The remaining events are attributed to three further
categories based on the multiplicity of $\cPqb$ tags
found in the event. We distinguish events with at least
two (``${\geq}2$ \cPqb\ tag''),
exactly one (``1 \cPqb\ tag''), and no $\cPqb$ tag (``0 \cPqb\ tag'').
These five categories are built separately in the muon and in the electron
channel leading to a total of ten categories.

\section{Mass reconstruction}
\label{sec:reco}
Hadronic jets, leptons, and $\ptvecmiss$ are used to reconstruct the
mass of the VLQ, denoted \mreco.
In signal events, the lepton in the final state always originates
from the decay of a $\PW$ boson, either the $\PW$ boson
from the VLQ decay or the $\PW$ boson from the top quark decay.
The neutrino four-momentum can thus be reconstructed from the
components of $\ptvecmiss$, the $\PW$ mass constraint, and the assumption
of mass\-less neutrinos.

In the case when a hadronic jet with a $\cPqt$ tag is found, \mreco is calculated
from the four-momentum of the $\cPqt$-tagged jet and the four-momentum of the
leptonically decaying $\PW$ boson. 
If several hadronic jets with $\cPqt$ tags are present,
the one with the largest angular distance to the reconstructed leptonic
$\PW$ boson decay is used. 
Once the $\cPqt$-tagged jet has been selected, all overlapping 
\akf\ jets in the event are removed in order to avoid double counting of energy.
For the shown \mreco distributions these events form the $\cPqt$ tag category.
For events in the other categories the hadronic part of the VLQ decay is reconstructed from
combinations of \akf$\mathrm{s}$ with $\abs{\eta} < 2.4$. 
Each possible jet assignment for the decays of the \PW\ boson
and \cPqt\ quark is tested exploiting the following $\chi^2$ quantity
\begin{linenomath}
\ifthenelse{\boolean{cms@external}}
{
\begin{flalign}
    \chi^2 &=
    \frac{\left(\mtop - \mtopmean\right)^2}{\sigma^2_{\cPqt}} +
    \frac{\left(\mW - \mWmean\right)^2}{\sigma^2_{\PW}} \nonumber \\
    &+ \frac{\left(\Delta R(\cPqt,\PW)-\pi \right)^2}{\sigma^2_{\Delta R}}+
    \frac{\left(\ptW / \ptt - 1\right)^2}{\sigma^2_{\pt}}.
   \label{eq:chi2}
\end{flalign}
}
{
\begin{flalign}
    \chi^2 =
    \frac{\left(\mtop - \mtopmean\right)^2}{\sigma^2_{\cPqt}} +
    \frac{\left(\mW - \mWmean\right)^2}{\sigma^2_{\PW}}
    + \frac{\left(\Delta R(\cPqt,\PW)-\pi \right)^2}{\sigma^2_{\Delta R}}+
    \frac{\left(\ptW / \ptt - 1\right)^2}{\sigma^2_{\pt}}.
   \label{eq:chi2}
\end{flalign}
}
\end{linenomath}
For each event, the jet assignment
with the maximum $\chi^2$ probability is selected.
For the $\chi^2$ quantity
the $\pt$ balance, $\ptW / \ptt$, the angular distance,
$\Delta R(\cPqt,\PW)$, and the reconstructed masses of the top quark candidate
$\mtop$ and the $\PW$ boson candidate $\mW$ are used.
The expected values $\mtopmean$ and $\mWmean$, and their standard
deviations $\sigma_{\cPqt}$ and $\sigma_{\PW}$ are obtained from
simulation for correctly reconstructed events
and it is verified that the values are independent
of the VLQ mass.
Here, correctly reconstructed events are defined by the assignment of jets to
generated $\cPqt$ quarks and $\PW$ bosons, where the generated particles from the
VLQ decay are unambiguously matched within a distance of $\Delta R < 0.4$ to the
reconstructed particles.
It was also verified in simulation that the expected values
of $\Delta R(\cPqt,\PW)$ and the $\pt$ balance are $\pi$ and $1$,
with their standard deviations $\sigma_{\Delta R}$ and $\sigma_{\pt}$.
In order to account for cases where the $\PW$ boson from the VLQ decay
decays into a lepton and neutrino, the $\chi^2$ is calculated for each permutation
with the second term omitted.
Cases where the hadronic decay products of the $\PW$ bosons or the top quark are
reconstructed in a single \akf\ are included
by omitting the first or second term in the calculation
of the $\chi^2$.

\begin{figure}
 \centering
\includegraphics[width=\cmsFigWidth]{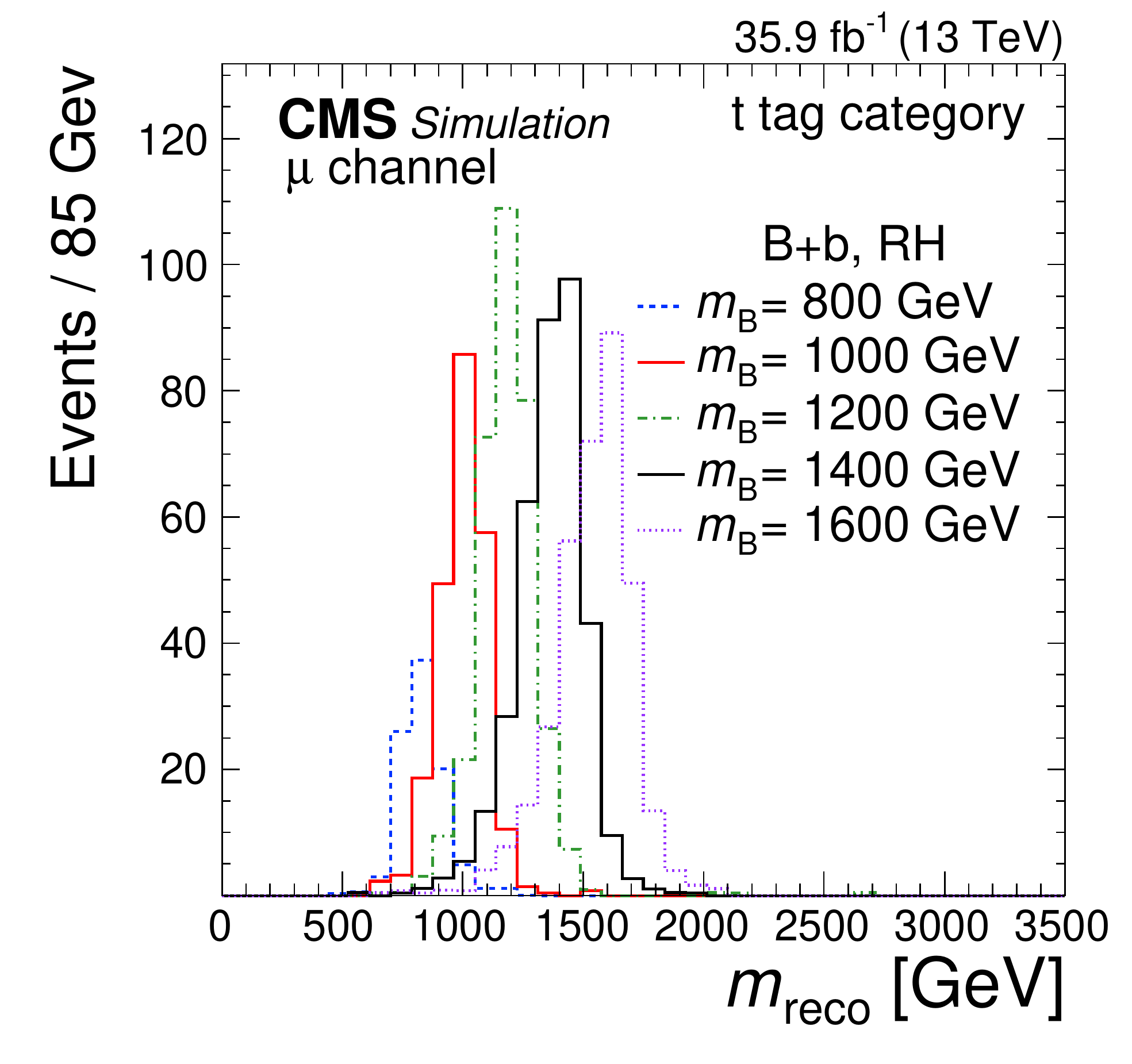}
\includegraphics[width=\cmsFigWidth]{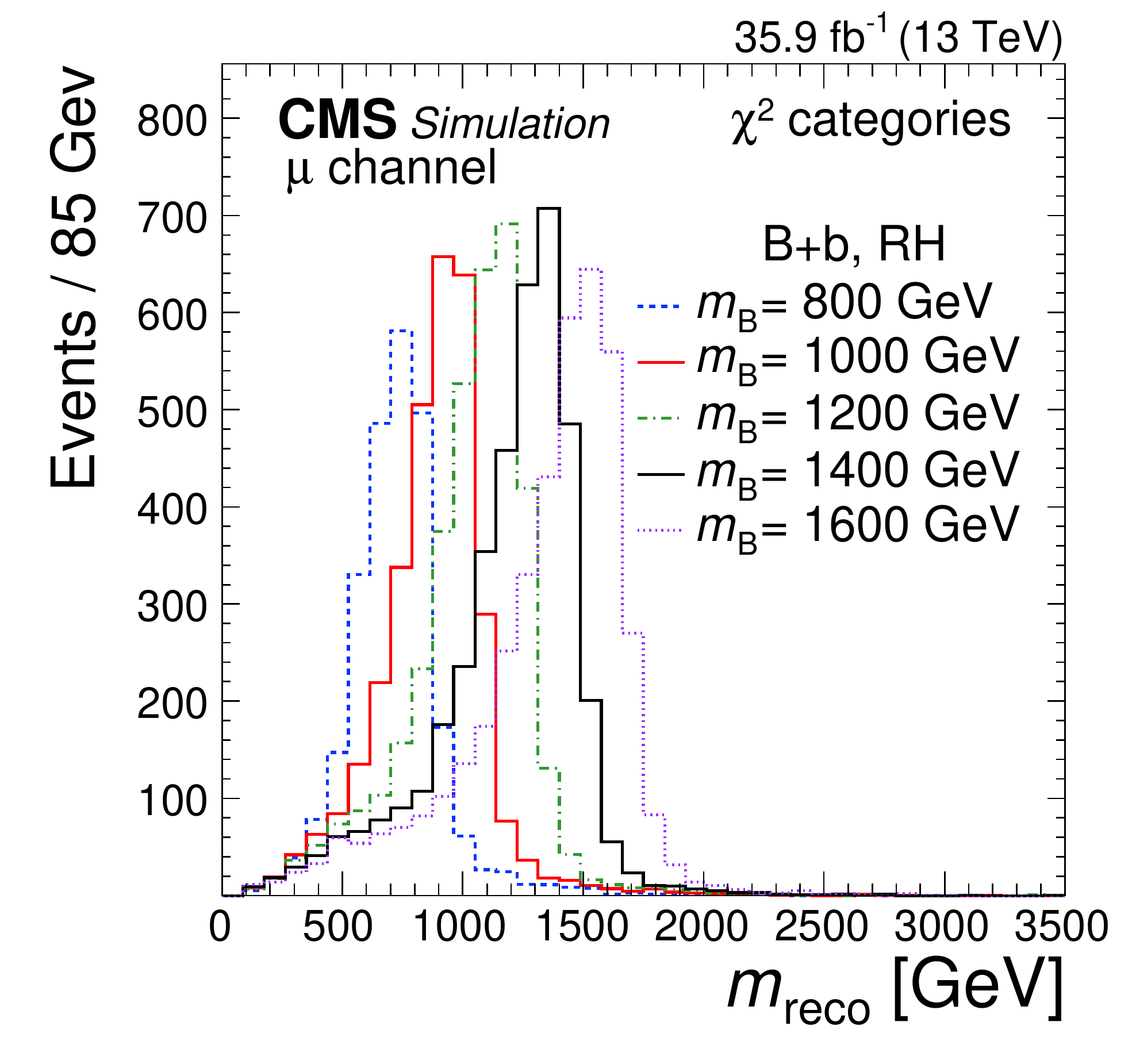}
 \caption{Distributions of $\mreco$ for the \PQB{}+\cPqb\ production
 mode, obtained for simulated events with a muon in the final state, reconstructed with a $\cPqt$ tag
 (\cmsLeft) and with the $\chi^2$ method (\cmsRight) for right-handed VLQ couplings and various VLQ masses $m_{\PQB}$. Signal events are shown assuming a production cross section of 1\pb and a relative VLQ decay width of 1\%.}
 \label{fig:Bt_mass_moun_rh}
\end{figure}

The distributions of \mreco in simulation for the \PQB{}+\cPqb\ production mode with right-handed couplings are shown in Fig.~\ref{fig:Bt_mass_moun_rh}
for events with a muon in the final state. The reconstruction
of events with a $\cPqt$ tag (\cmsLeft) is best suited for high VLQ masses
where the decay products of the top quark are highly boosted, while
the $\chi^2$ method (\cmsRight) yields a stable performance for all VLQ masses,
where the decay products of the $\PW$ boson and top quark are
reconstructed from several jets. Additionally, the latter method
enables the reconstruction of events with a lepton from the top
quark decay chain.
Mass resolutions between 10--15\% are achieved for both reconstruction methods,
with peak values of the $\mreco$ distributions at the expected values.
The VLQs with left-handed couplings (not shown) have a lower selection
efficiency by 20--25\% because of a smaller lepton $\pt$, on average,
but otherwise features a behaviour similar to VLQs with right-handed couplings.
Distributions obtained for the final states with an electron are similar
to those with a muon.

\section{Background estimation}
\label{sec:background}
The data sample obtained after the selection is then divided into
a signal region with a jet in the forward region of the detector with
$2.4 < \abs{\eta} < 4.0$ and a control region without such a jet.
The distribution of background processes in the signal region
is estimated using the shape of the \mreco\ distribution in the control region.
Residual differences in the shapes of the \mreco distributions
between signal and control regions are investigated in each of the
signal categories by using simulated SM events.
Differences can arise from different
background compositions in signal and control regions due to the
presence of a forward jet.
The observed differences are small, with average values of 10\%,
and are corrected for by multiplicative factors applied to the background predictions
in the validation and signal regions. The largest differences are observed for \mreco
values below 800\GeV, with values no larger than about 20\%.

\begin{figure*}
 \centering
 \includegraphics[width=0.45\textwidth]{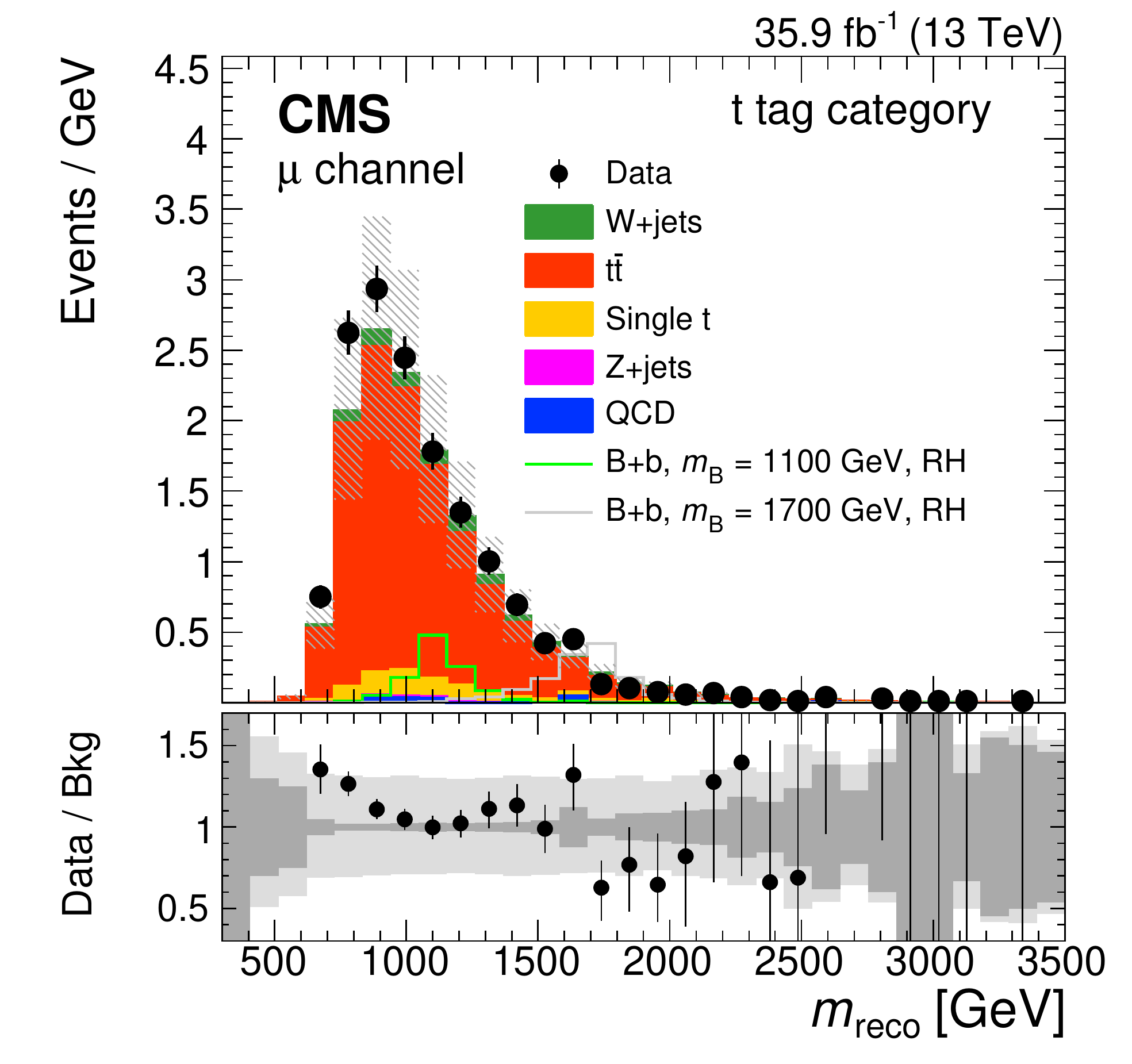}	
 \includegraphics[width=0.45\textwidth]{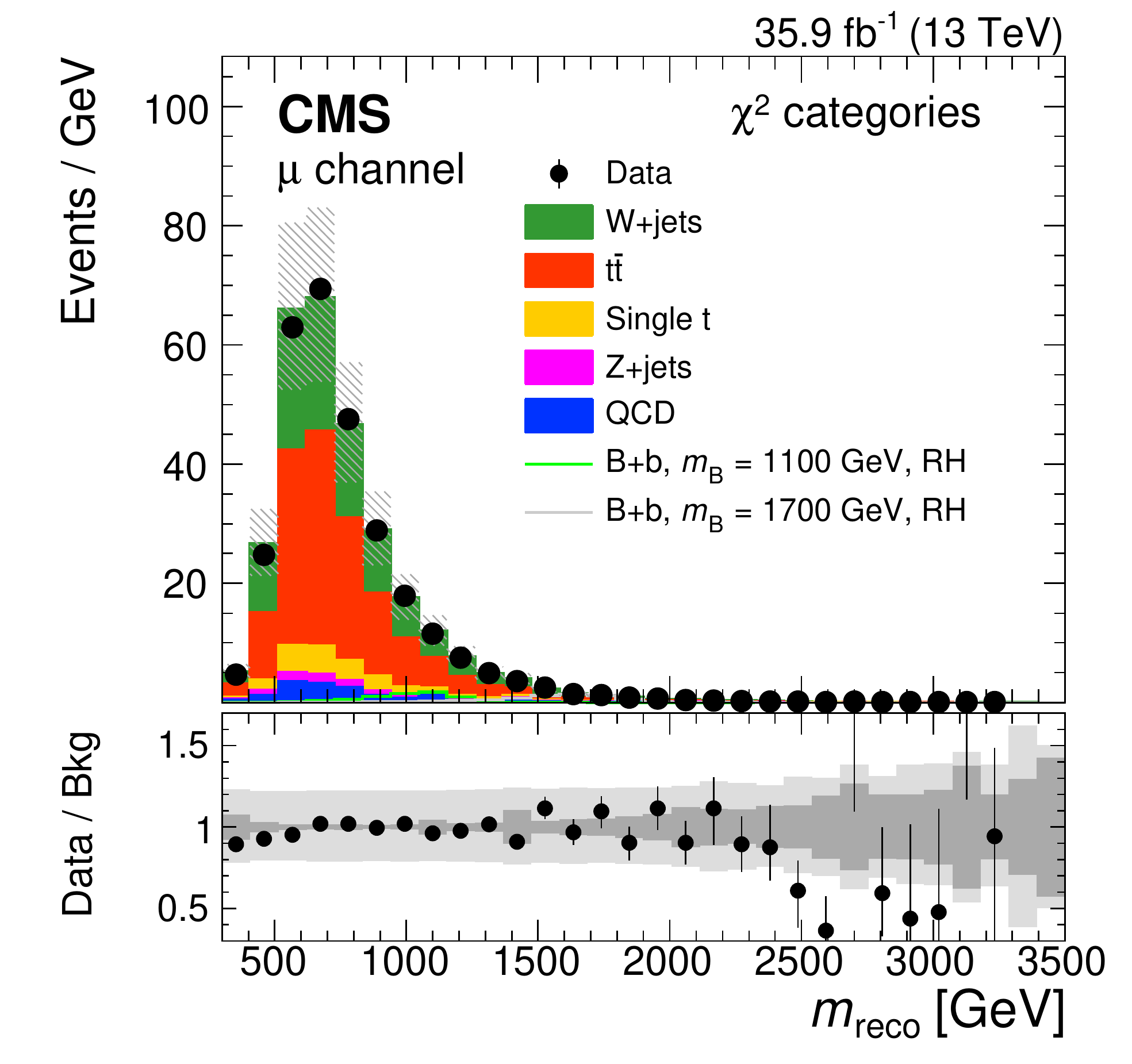}
 \includegraphics[width=0.45\textwidth]{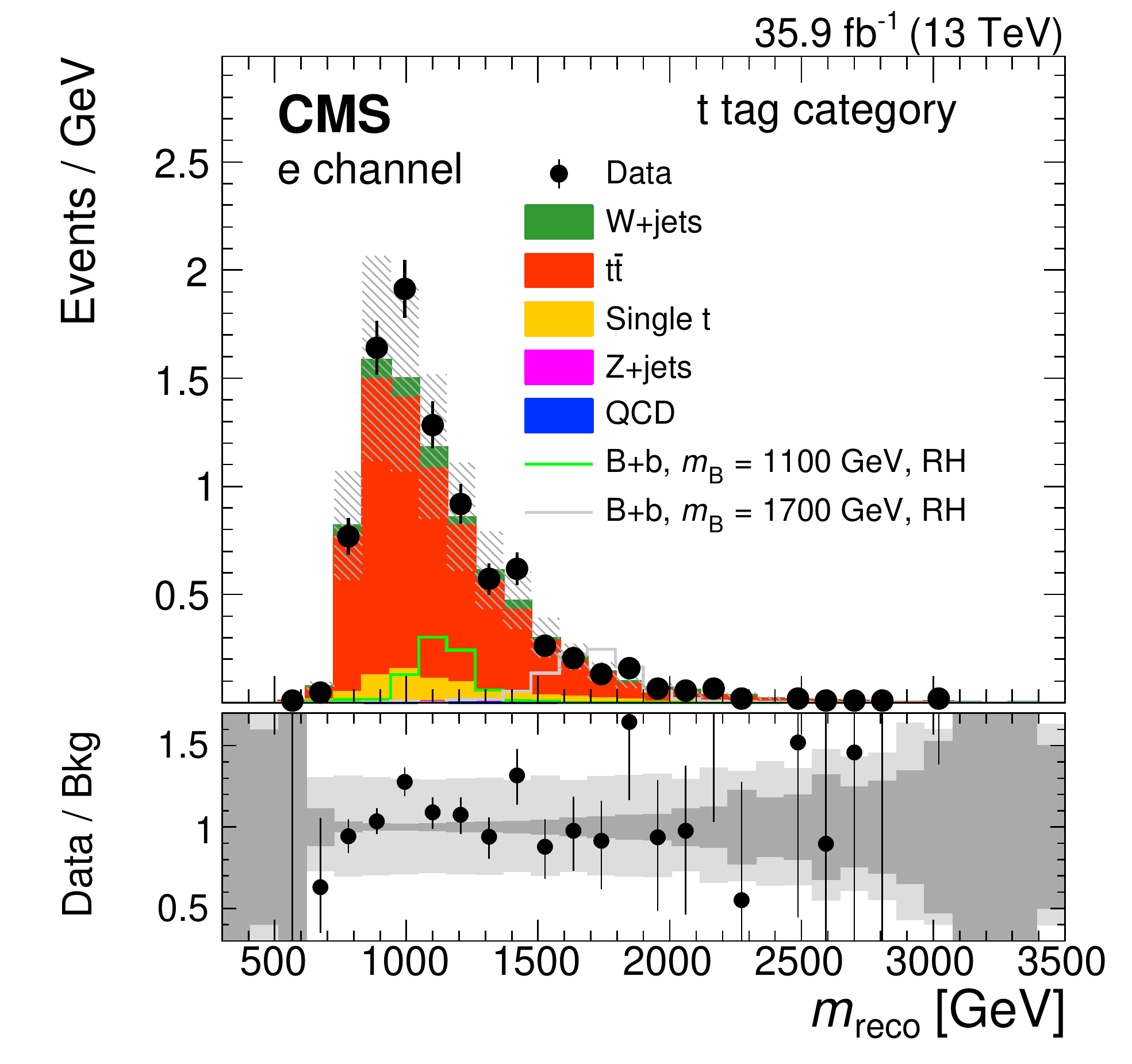}
 \includegraphics[width=0.45\textwidth]{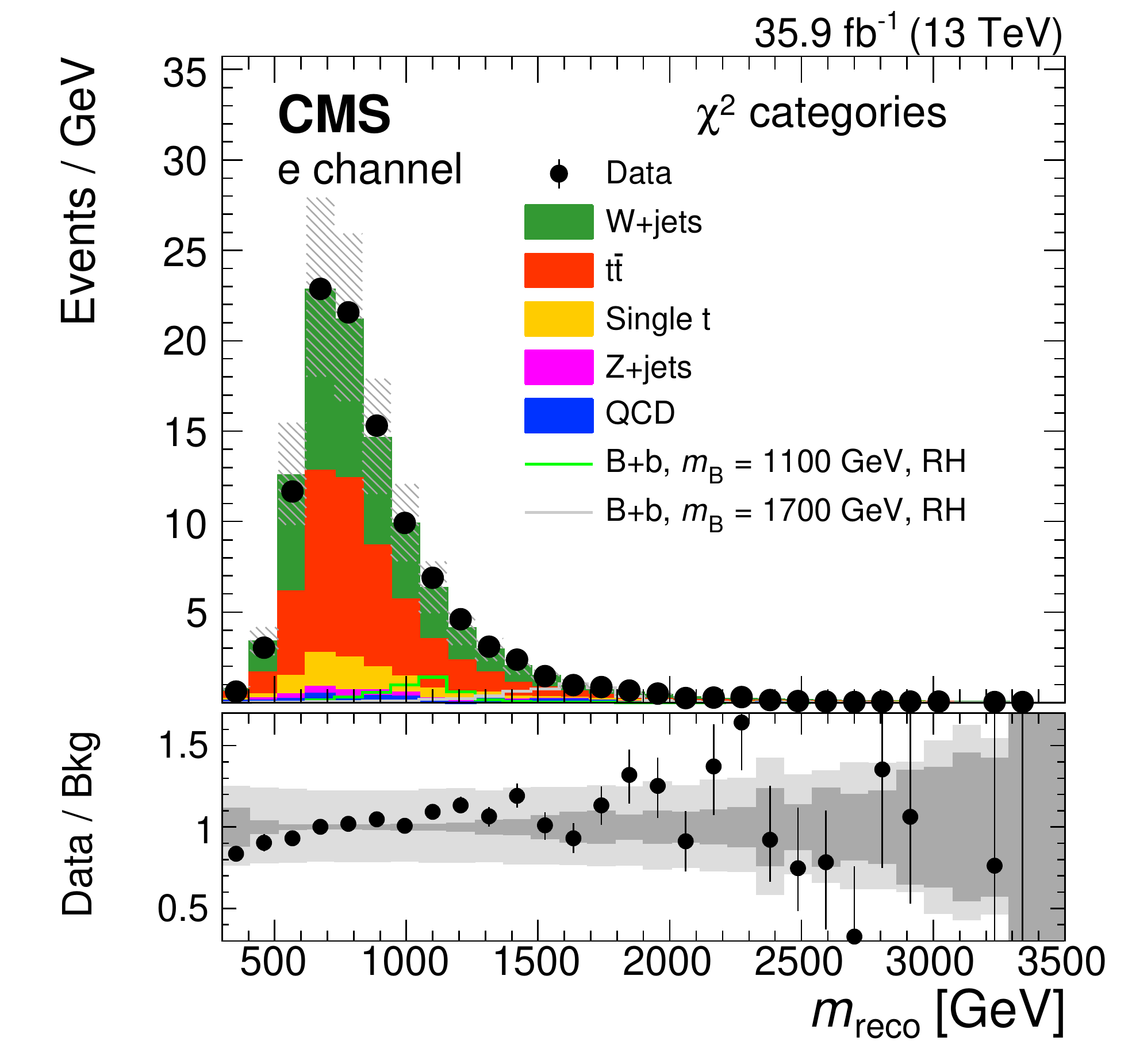}
 \caption{Distributions of \mreco in data and simulation in the control
 region for the muon (upper) and electron (lower) channels
 for events reconstructed with a $\cPqt$ tag (left) and with
 the $\chi^2$ method (right). The VLQ signal is shown for the \PQB{}+\cPqb\ production mode and right-handed VLQ couplings.
 The vertical bars illustrate the statistical
 uncertainties on the data, while the shaded area shows the total
 uncertainties for the background simulation. The lower panels show the
 ratio of data to the background prediction. The dark and light gray bands
 correspond to the statistical and total uncertainties,
 respectively.
 \label{fig:background_mu} }
\end{figure*}

In order to validate the VLQ mass reconstruction, data are compared to
simulation in the control region. In Fig.~\ref{fig:background_mu}
the distributions of \mreco are shown
in the muon (upper) and electron (lower) channels for events with
a $\cPqt$ tag (left) and events reconstructed with the $\chi^2$
method (right). The \ttbar and $\cPqt\PW$ standard model processes provide
irreducible backgrounds in the reconstructed VLQ mass distributions, showing good agreement between the data and simulation.
The contribution of signal events in the control region  is small and is taken into account by a simultaneous fit to signal and control regions in the statistical extraction of the results.

\begin{figure*}
 \centering
 \includegraphics[width=0.45\textwidth]{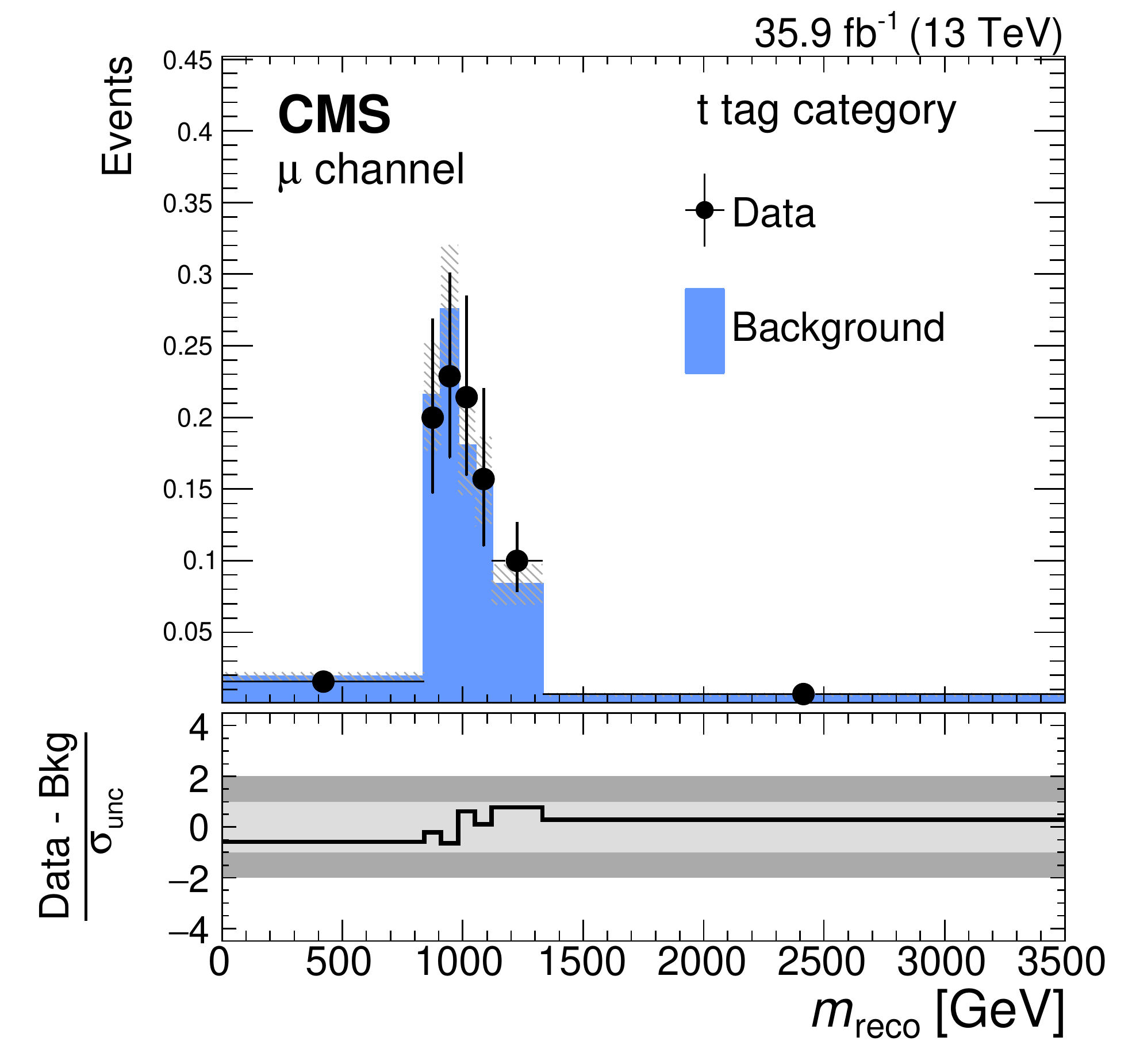}
 \includegraphics[width=0.45\textwidth]{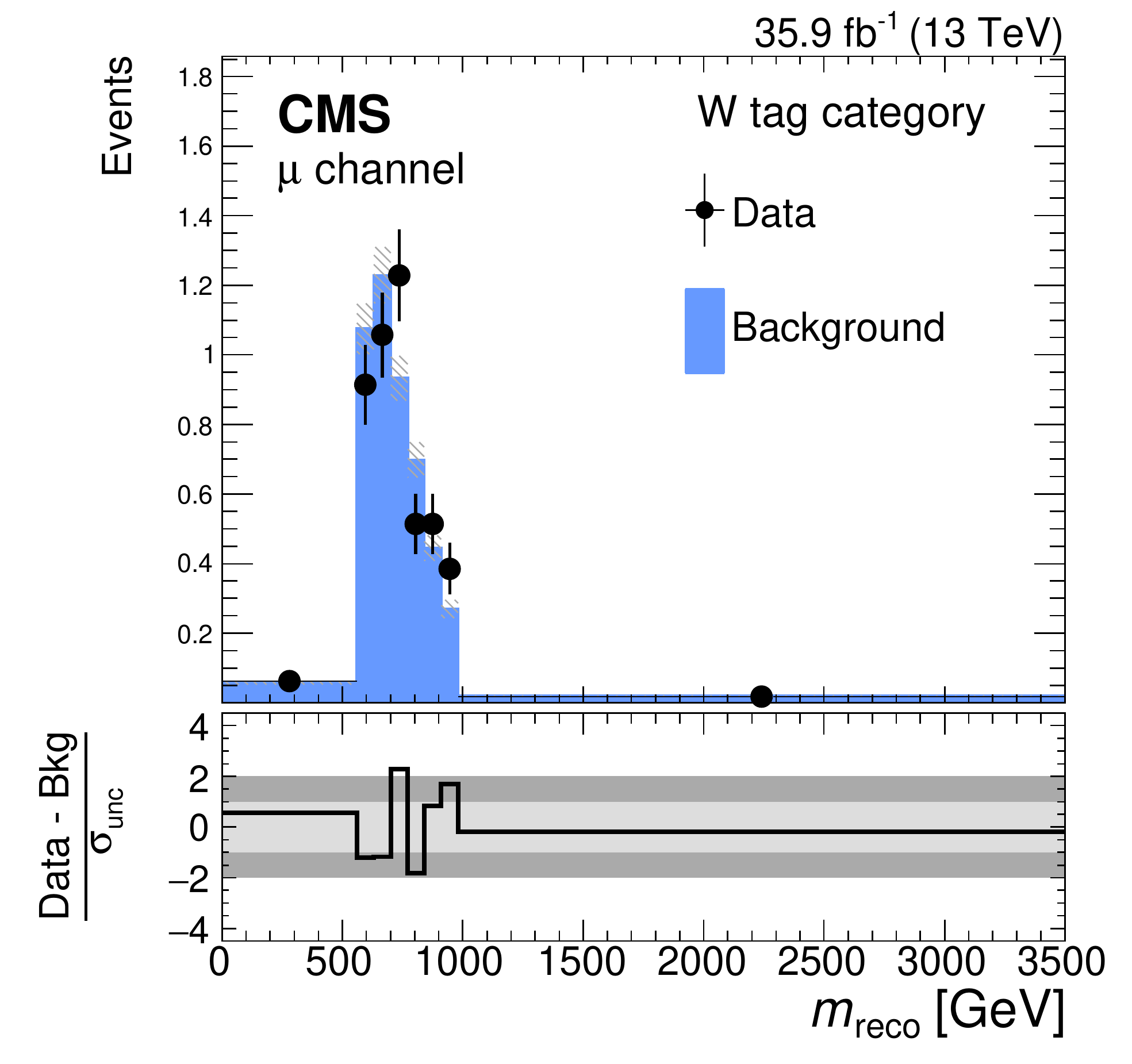}
 \includegraphics[width=0.45\textwidth]{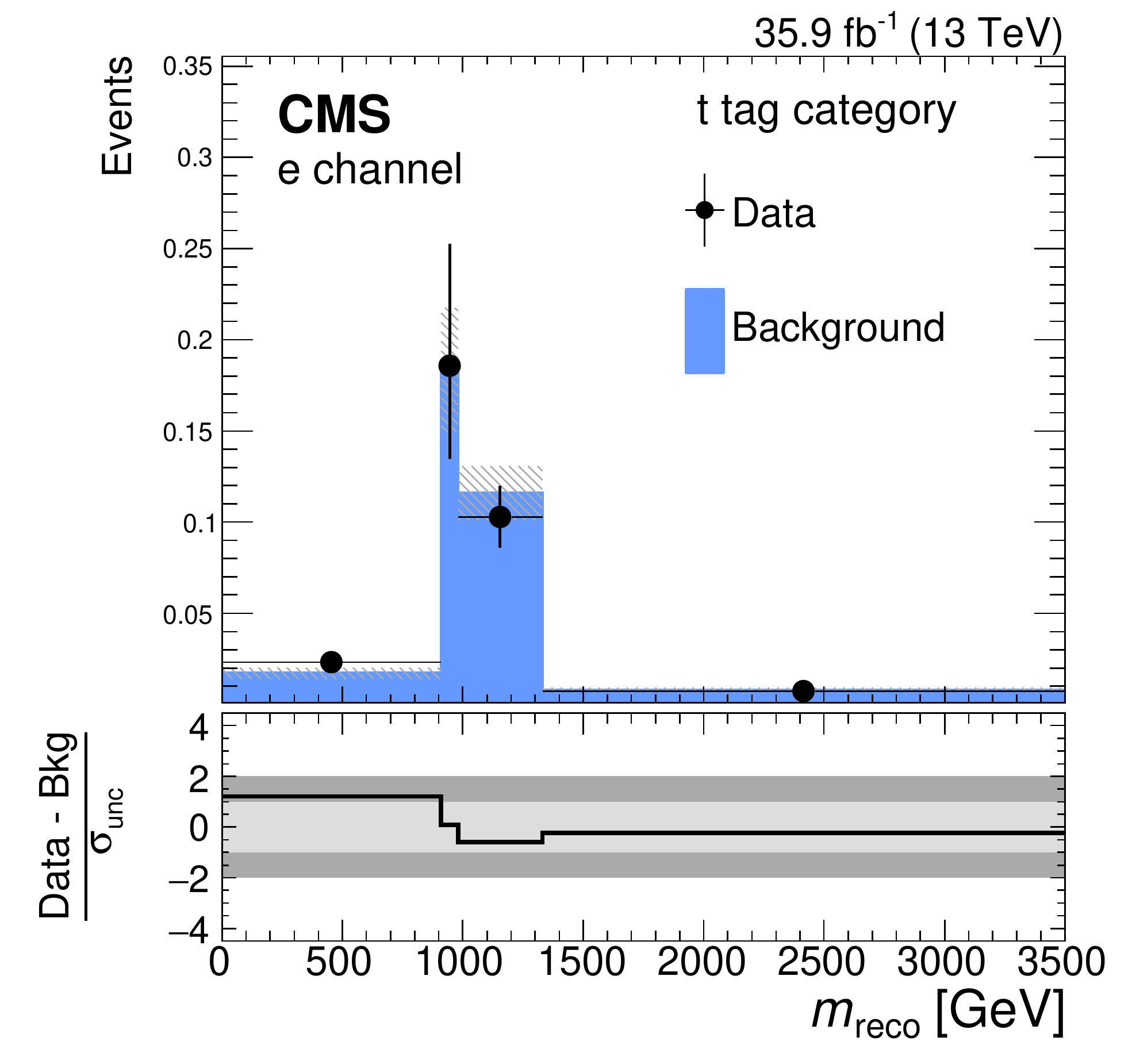}
 \includegraphics[width=0.45\textwidth]{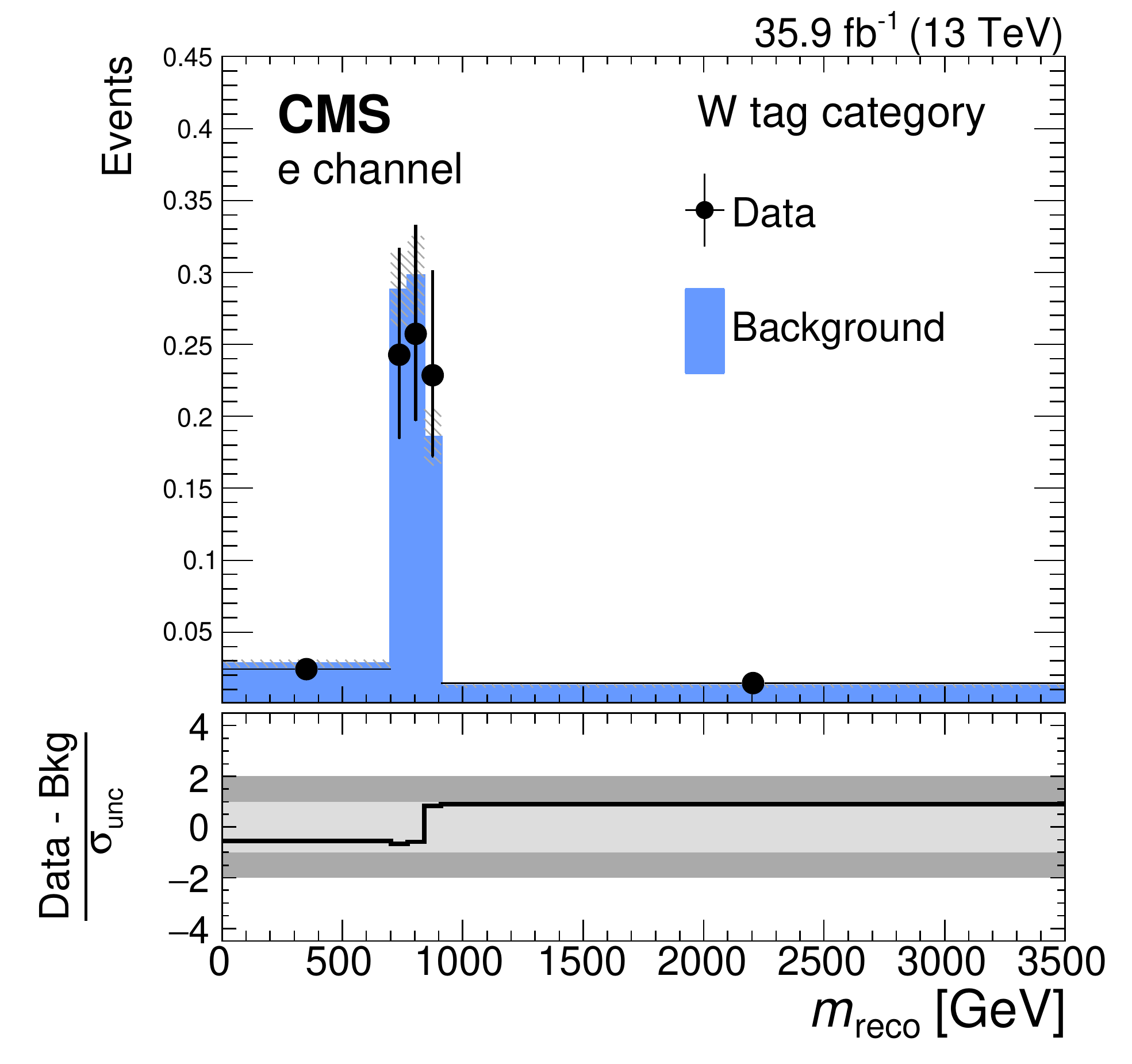}
 \caption{Distributions of \mreco in the validation region
 of the two most sensitive categories in the muon channel (upper)
 and electron channel (lower). The lower panels show the difference of data
 and background expectations in units of the total (stat. and sys.)
 uncertainty on the background estimate.
 \label{fig:validation_mu}}
\end{figure*}

In order to validate the background estimation, a validation region
is constructed from requiring events with reconstruction
$p$-values smaller than 0.08. The $p$-values are calculated as
the probability of obtaining the $\chi^2$ as given by Eq.~\eqref{eq:chi2},
where the number of degrees of freedom of the selected hypothesis
are taken into account. For events with a $\cPqt$ tag, the same $\chi^2$
quantity is evaluated for the selected hypothesis.
The validation region has an order of magnitude fewer events than the
signal region and a negligible amount of signal contamination.
The \mreco distributions for the two most sensitive categories are
shown in Fig.~\ref{fig:validation_mu} for the muon (upper) and
electron (lower) channels.
The observed number of events is found to be in good
agreement with the predicted number of events from the background estimation
in the validation region, with no statistically significant deviations.
Similar observations are made for the other signal categories.

\section{Systematic uncertainties}
\label{sec:systematic}
Systematic uncertainties can affect both the overall normalisation
of background components and the shapes of the \mreco distributions
for signal and background processes.
The main uncertainty in the shape of the \mreco distribution from the
background estimation based on a control region in data is related to the kinematic difference
between the signal and control regions. Correction factors are applied
to account for this difference, obtained from SM simulations. 
These uncertainties have a size of 10\% on average, with maximum values 
of 20\% at small values of \mreco. Compared to these uncertainties, the 
effects from uncertainties in the SM simulations are negligible on the background estimation, 
as these cancel to a large degree when building the ratios between signal and control regions.
The uncertainties in the overall
normalisation of the background predictions are obtained from a fit
to the data in the signal region.

Uncertainties in the MC simulation are applied to all simulated
signal events. In the following, the systematic uncertainties are summarized.

\begin{itemize}

 \item The uncertainty in the integrated luminosity measurement
 recorded with the CMS detector in the 2016 run
 at $\sqrt{s}=13\TeV$ is 2.5\%~\cite{CMS-PAS-LUM-17-001}.

 \item The estimation of pileup effects is based on the total
 inelastic cross section. This cross section is determined to
 be 69.2\unit{mb}. The uncertainty is taken into account by
 varying the total inelastic cross section by 4.6\%~\cite{Chatrchyan:2012nj}.

 \item Simulated events are corrected for lepton identification,
 trigger, and isolation efficiencies. The corresponding
 corrections are applied as functions of $\abs{\eta}$ and $\pt$.
 The systematic uncertainties due to these corrections are
 taken into account by varying each correction factor within its uncertainty.

 \item The scale factors for the jet energy scale
 and resolution  are determined as functions of $\abs{\eta}$ and $\pt$~\cite{Khachatryan:2016kdb}.
 The effect of the uncertainties in these scale factors are considered by varying
 the scale factors within their uncertainties. Jets with distance parameters of
 0.4 and 0.8 are modified simultaneously. The results of variations for \akf$\mathrm{s}$ are propagated to the measurement of \ptvecmiss.

 \item The uncertainties due to the PDFs are evaluated by considering
 100 replicas of the NNPDF~3.0 set according to the procedure
 described in Ref.~\cite{Butterworth:2015oua}.
 The associated PDF uncertainties in the signal acceptance are estimated following the prescription for the LHC~\cite{Butterworth:2015oua}.

 \item Uncertainties associated with variations of the factorisation $\mu_\mathrm{f}$ and renormalisation scales $\mu_\mathrm{r}$
 are evaluated by varying the respective scales independently, by factors of 0.5 and 2.

 \item Corrections for the \cPqb\ tagging efficiencies and
 misidentification rates for \akf$\mathrm{s}$, and subjets of \ake$\mathrm{s}$ are applied.
 These are measured as a function of the jet $\pt$~\cite{Sirunyan:2017ezt}.
 The corresponding uncertainties are taken into account by varying
 the corrections within their uncertainties for heavy- and light-flavour
 jets separately.

 \item An uncertainty on the \cPqt\ tagging efficiency
 of $+7$ and $-4\%$ is applied to signal events with a \cPqt\ tag~\cite{CMS-PAS-JME-16-003}.
 The uncertainty on the $\PW$ tagging efficiency is determined
 from jet mass resolution (JMR) and scale (JMS) uncertainties,
 which are added in quadrature. An additional JMR uncertainty is derived from
 the differences in the hadronisation and shower models of
 \PYTHIA and \HERWIGpp~\cite{Bahr:2008pv}. The uncertainty depends on the \pt
 of the \PW\ boson; for VLQs with a mass of 700\GeV it is around 2\% and for
 a mass of 1800\GeV it is around 6\%. An uncertainty of 1\% is assigned to the JMS, 
 as obtained from studies of the jet mass in fully merged hadronic \PW\ boson decays.

\end{itemize}

In Table~\ref{tab:unc_impact}, a summary of the uncertainties
considered for signal events is shown, where the largest uncertainties
come from the jet energy scale and the jet tagging. For the uncertainties
connected to the PDF, $\mu_\mathrm{f}$ and $\mu_\mathrm{r}$ only the signal
acceptance and shape differences are propagated. The uncertainties with the
largest impact on the analysis are the uncertainties associated with
the data-driven background estimation, being more than two times larger than
the jet energy scale uncertainties in the signal.

\begin{table}
  \centering
  \topcaption{Uncertainties considered for simulated signal
  events in the \PQB{}+\cPqb\ production mode ($m_{\PQB{}} = 900 \GeV$) for right-handed VLQ couplings for the \cPqt\ tag and \PW\ tag categories. The uncertainties in the \cPqb\ tag categories are of comparable size to those in the \PW\ tag category.}
  \begin{tabular}{l c  c  c  c}

     Uncertainty            &        & $\cPqt$ tag  [$\%$] & \PW\ tag  [$\%$]  \\ \hline
     \PW\ tagging           & Rate   & \NA               & 3.3          \\
     \cPqt\ tagging         & Rate   & $^{+7}_{-4}$      & \NA          \\
     Luminosity             & Rate   &  2.5              & 2.5          \\
     Pileup                 & Shape  & 1--3              & 0.2          \\
     Lepton reconstruction  & Shape  & 2--3              & 2--3         \\
     \cPqb\ tagging         & Shape  & 2.5               & 2.5          \\
     Jet energy scale       & Shape  & 2--6              & 1--5         \\
     Jet energy resolution  & Shape  & 1--2              & 1--2         \\
     PDF                    & Shape  & 2--3              & 0.5          \\
     $\mu_f$ and $\mu_r$    & Shape  & 0.3               & 0.2          \\
  \end{tabular}
  \label{tab:unc_impact}
\end{table}

\section{Results}
\label{sec:results}
The \mreco distributions in the ten categories are measured in the signal
and control region, which are defined by the presence or absence of a
forward jet with $\abs{\eta} > 2.4$.
For the background estimate in the signal regions, a simultaneous
binned maximum likelihood fit of both regions is performed
using the \textsc{Theta}~\cite{theta} package.
In these fits, the signal cross section and the background
normalisations in the different signal categories are free parameters.
The shapes of the \mreco distributions for the SM background in the signal regions are taken
from the corresponding control regions.
Systematic uncertainties are taken into account as additional
nuisance parameters. A common nuisance parameter is used for
uncertainties in the muon and electron channels if a similar effect is
expected on the shape or normalisation of the
\mreco distribution in both channels similarly.
The nuisance parameters for the shape uncertainties
are taken to be Gaussian distributed. For the uncertainties on the normalisation log-normal prior distributions are assumed.

\begin{figure*}
 \centering
 \includegraphics[width=0.45\textwidth]{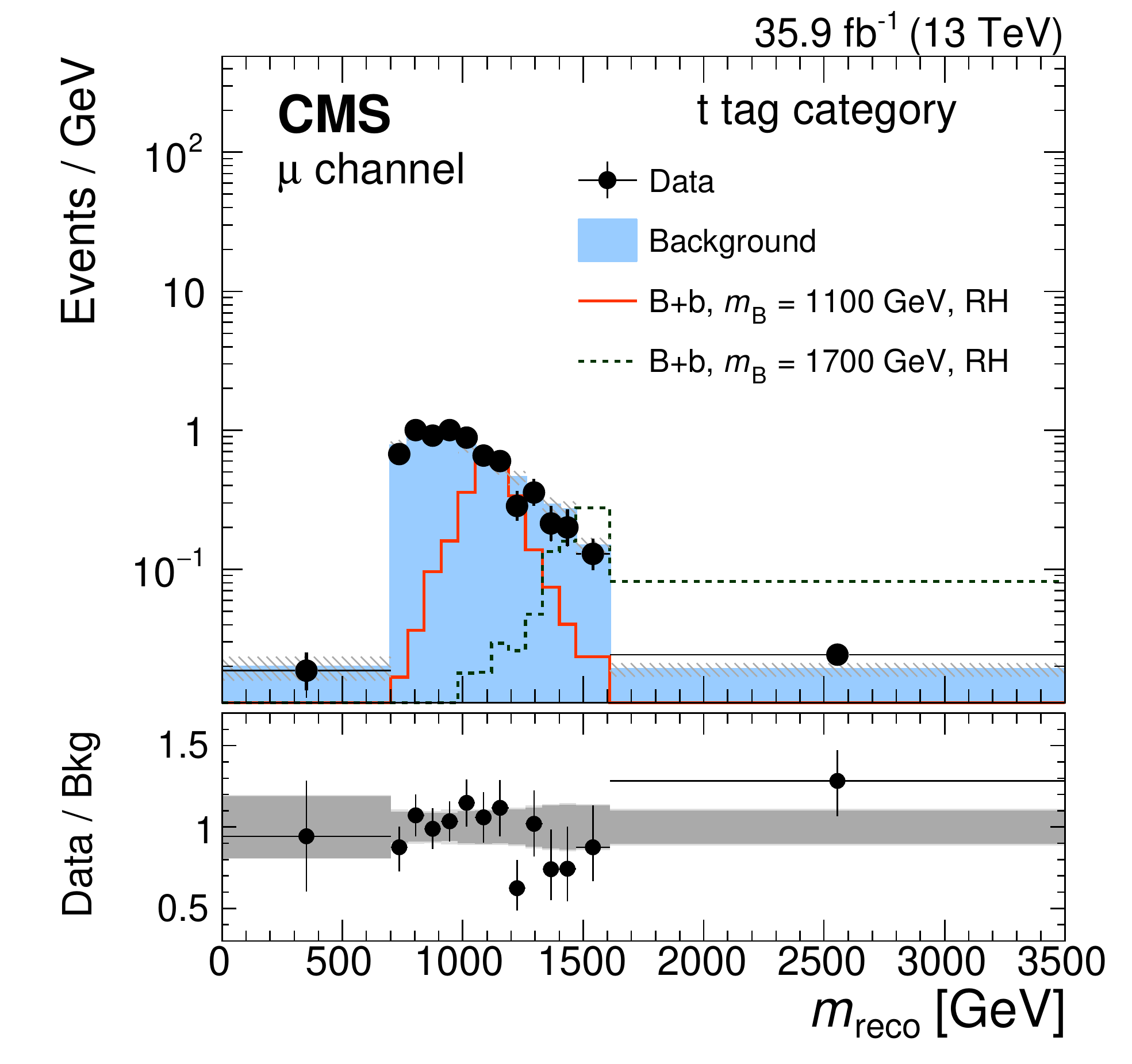}
 \includegraphics[width=0.45\textwidth]{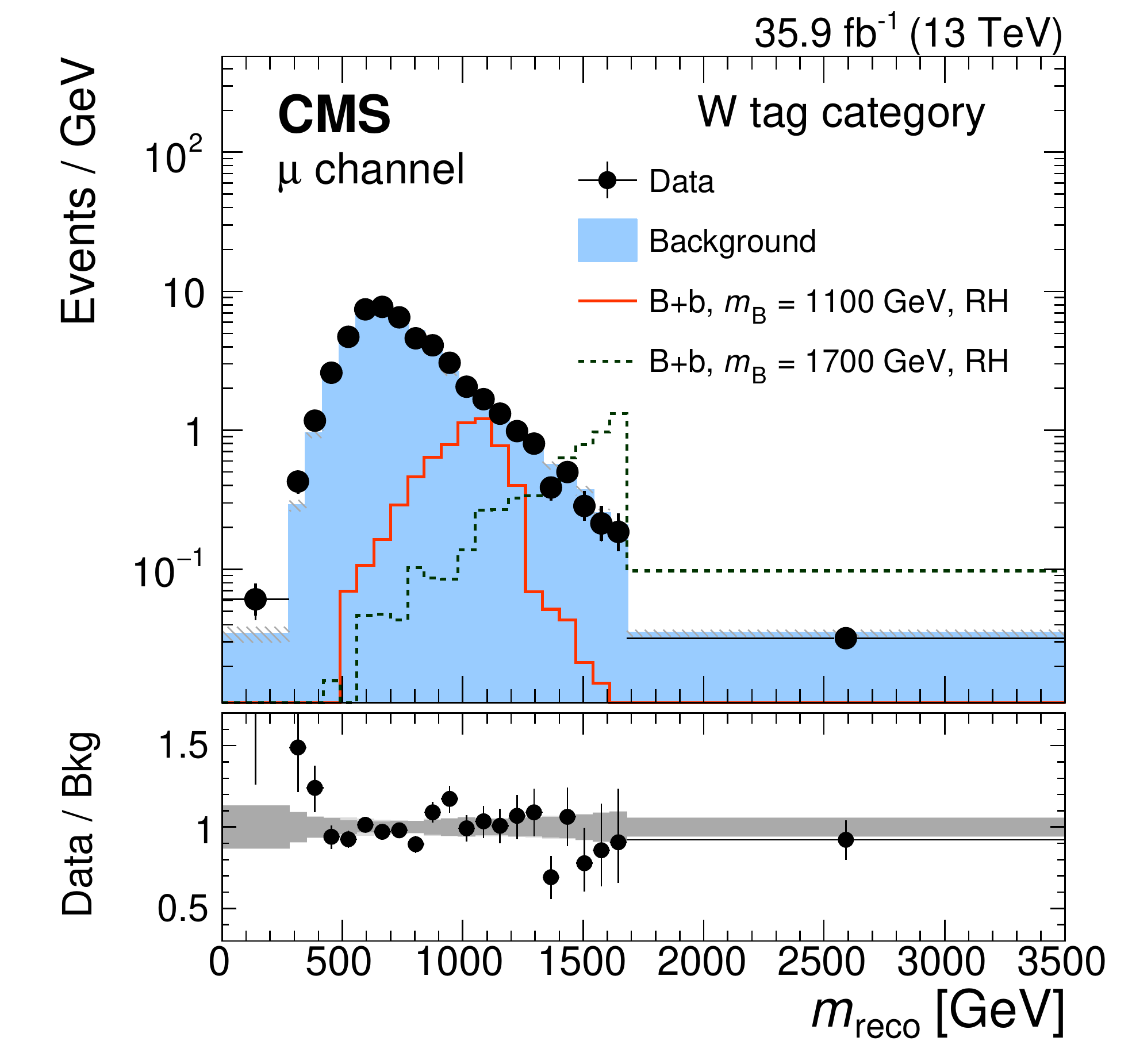}
 \includegraphics[width=0.45\textwidth]{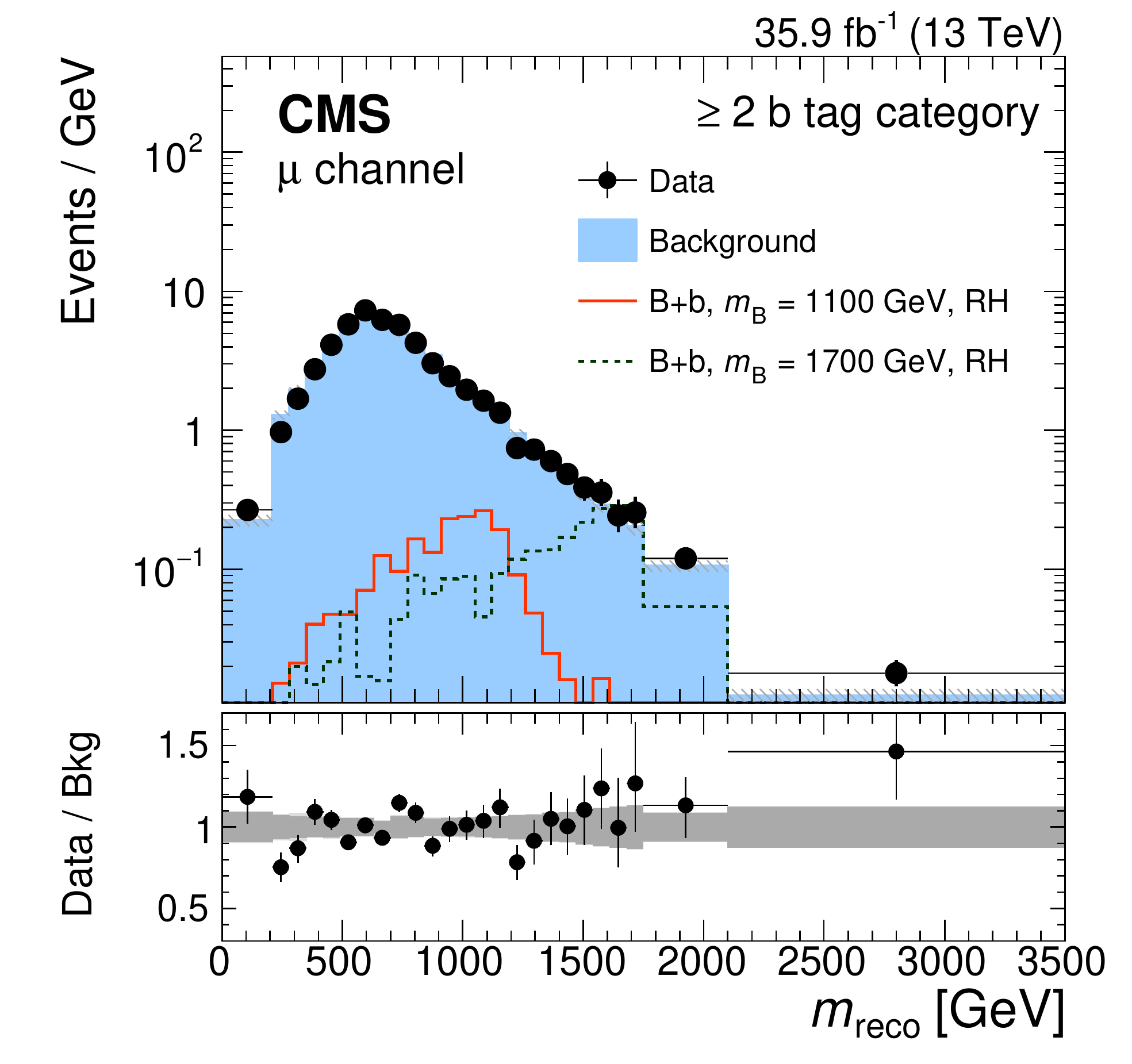}
 \includegraphics[width=0.45\textwidth]{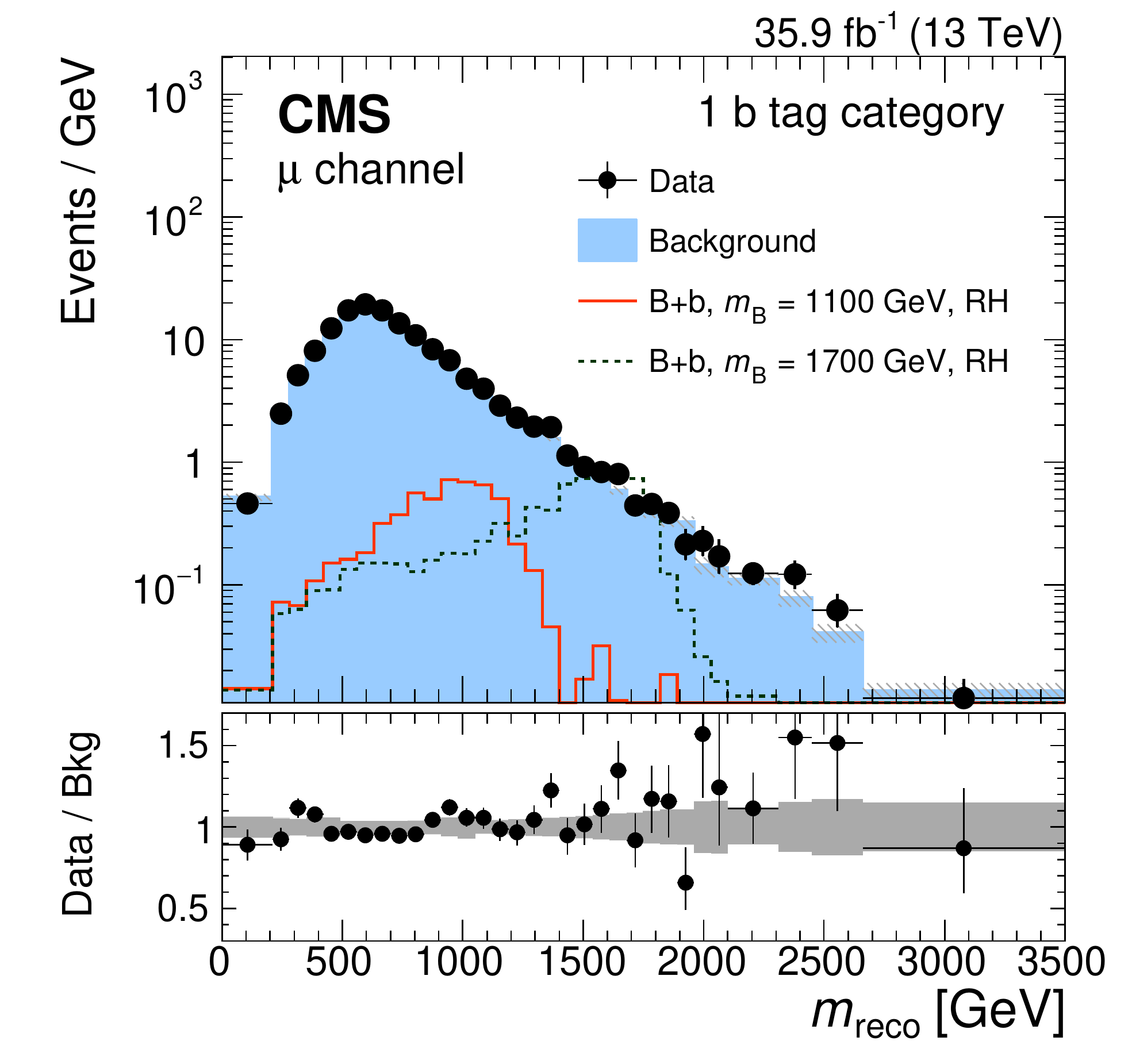}
 \includegraphics[width=0.45\textwidth]{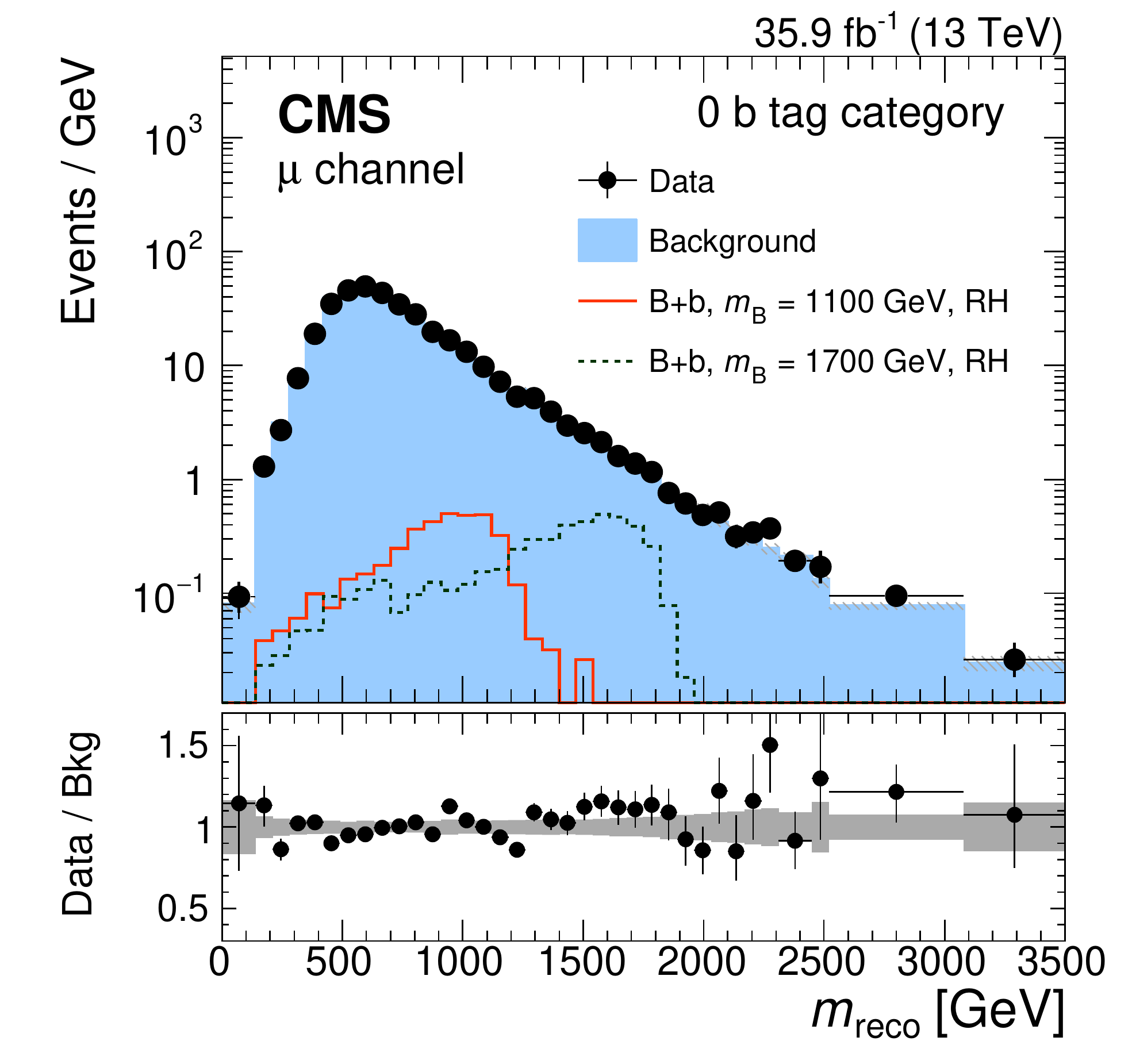}
 \caption{Distributions of \mreco\ measured in the signal region for
 events with a jet in the forward direction with $\abs{\eta}>2.4$ in the muon channel.
 Shown are the sensitive categories:
 \cPqt\ tag (upper left),
 \PW\ tag (upper right),
 ${\geq}2$ \cPqb\ tag (middle left),
 1 \cPqb\ tag (middle right) and 0 \cPqb\ tag (lower).
 The background prediction is obtained from control regions as
 detailed in the main text. The distributions from two example signal samples
 for the \PQB{}+\cPqb\ production mode with right-handed VLQ couplings
 with a cross section of 1\pb and a relative width of 1\% are shown for illustration.
 \label{fig:signalregion_mu} }
\end{figure*}

\begin{figure*}
 \centering
 \includegraphics[width=0.45\textwidth]{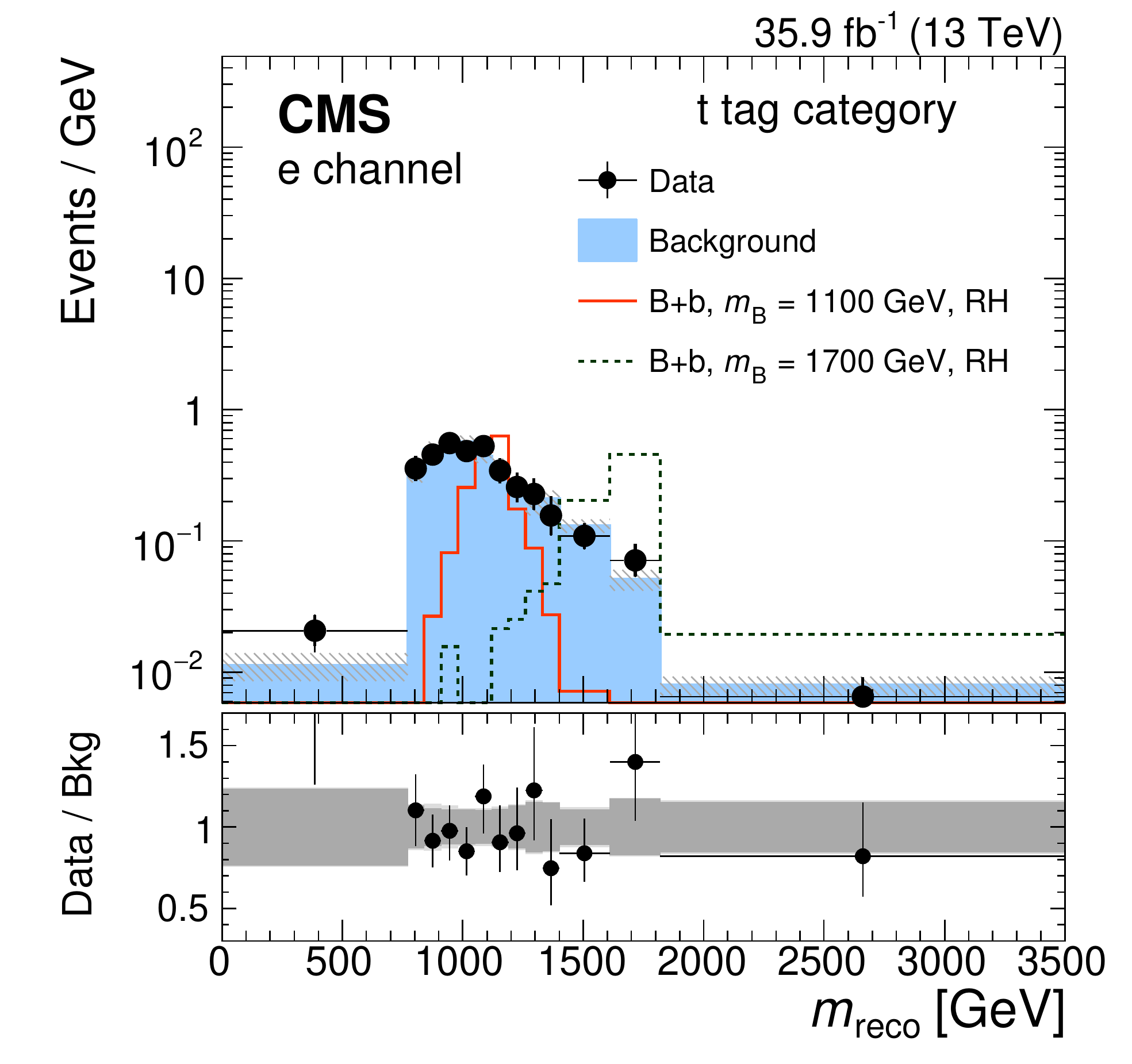}
 \includegraphics[width=0.45\textwidth]{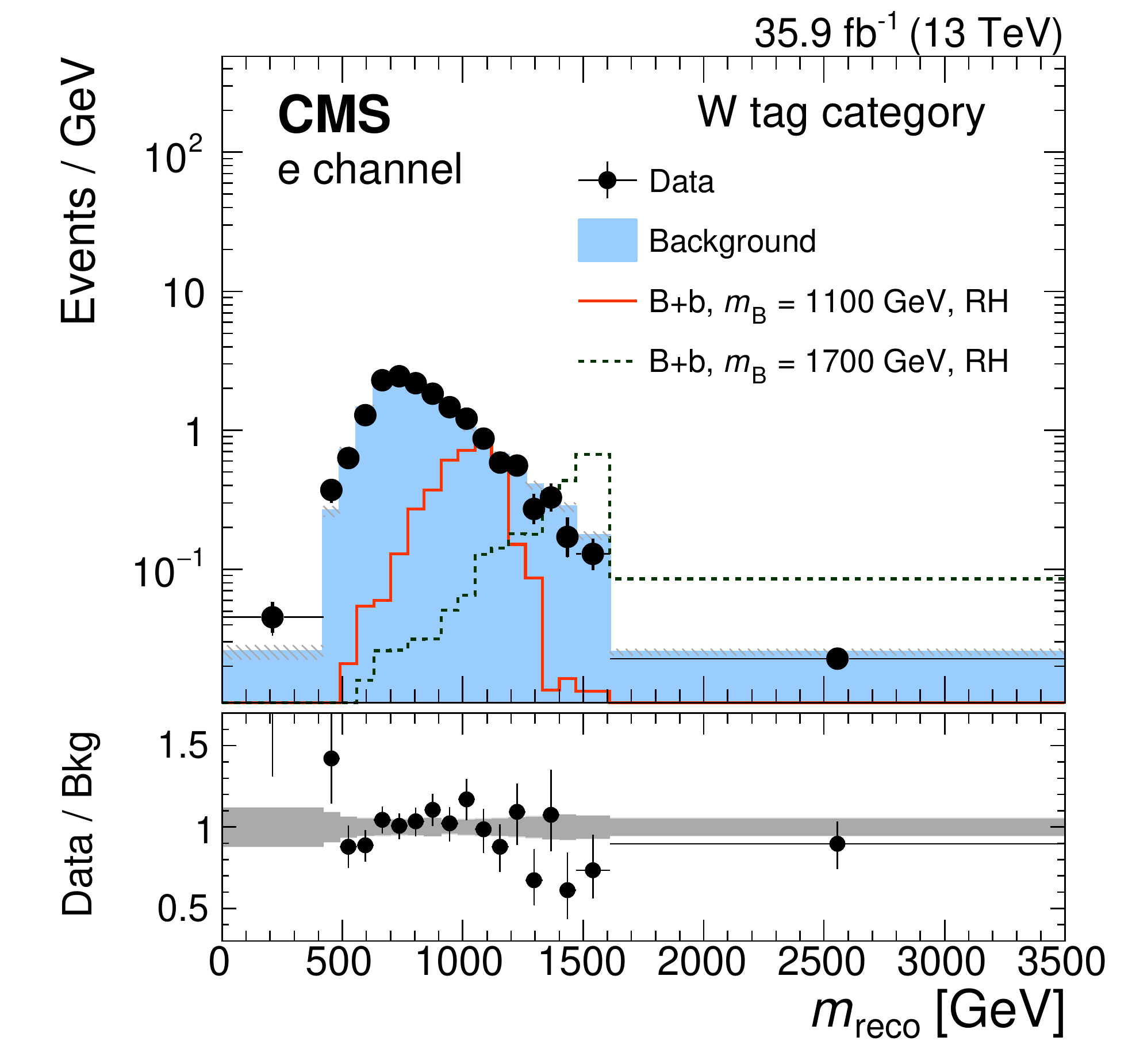}
 \includegraphics[width=0.45\textwidth]{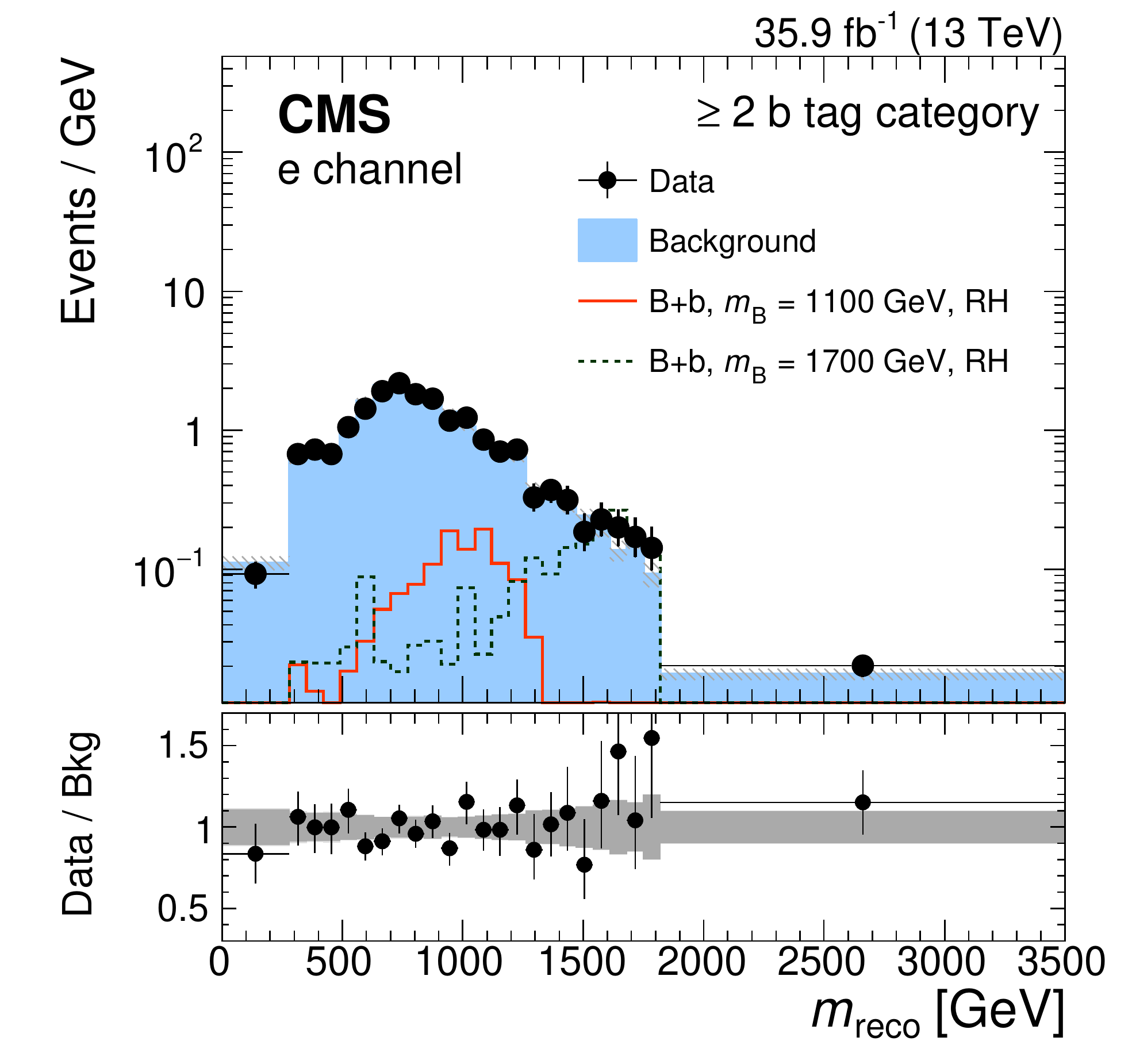}
 \includegraphics[width=0.45\textwidth]{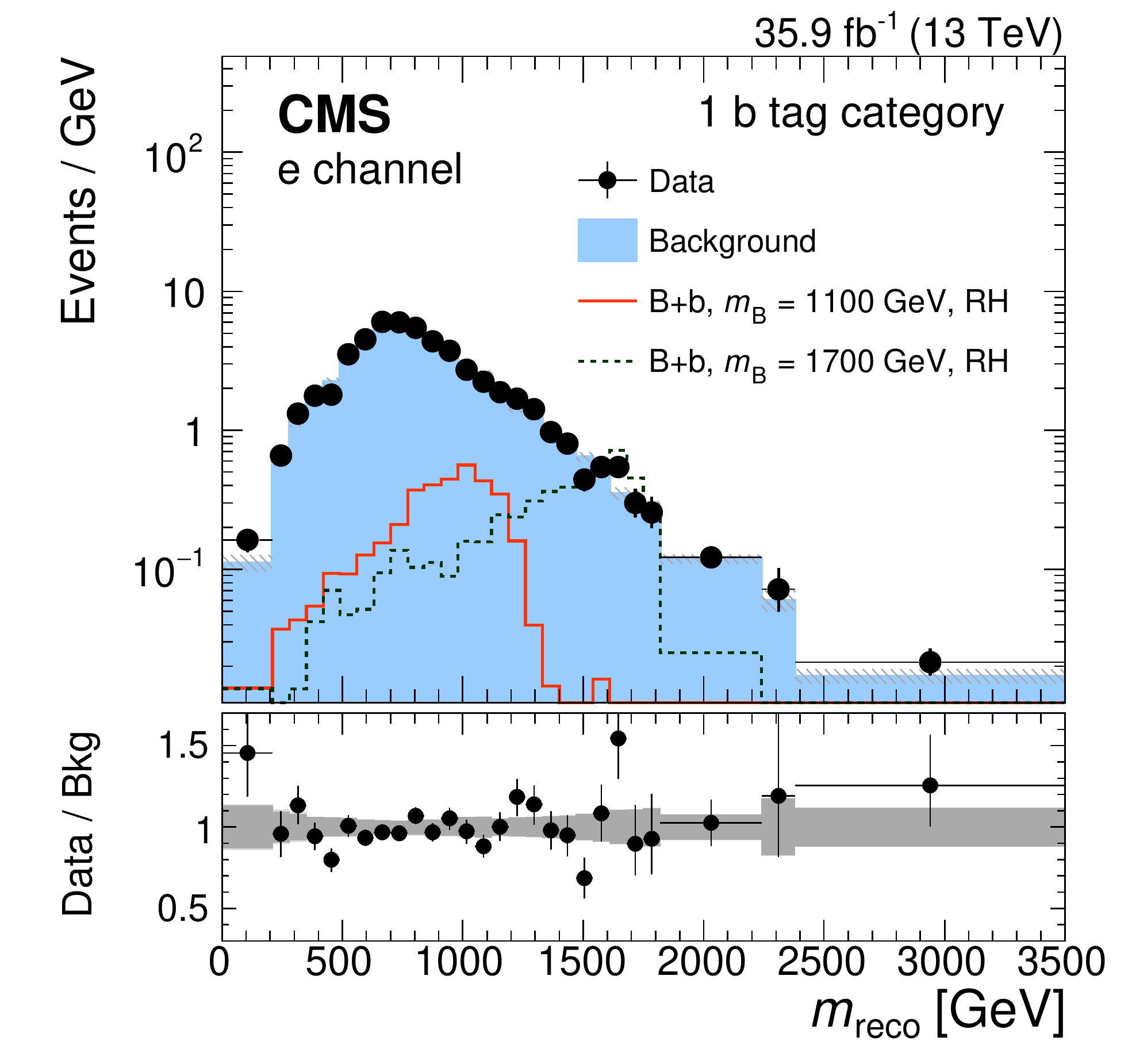}
 \includegraphics[width=0.45\textwidth]{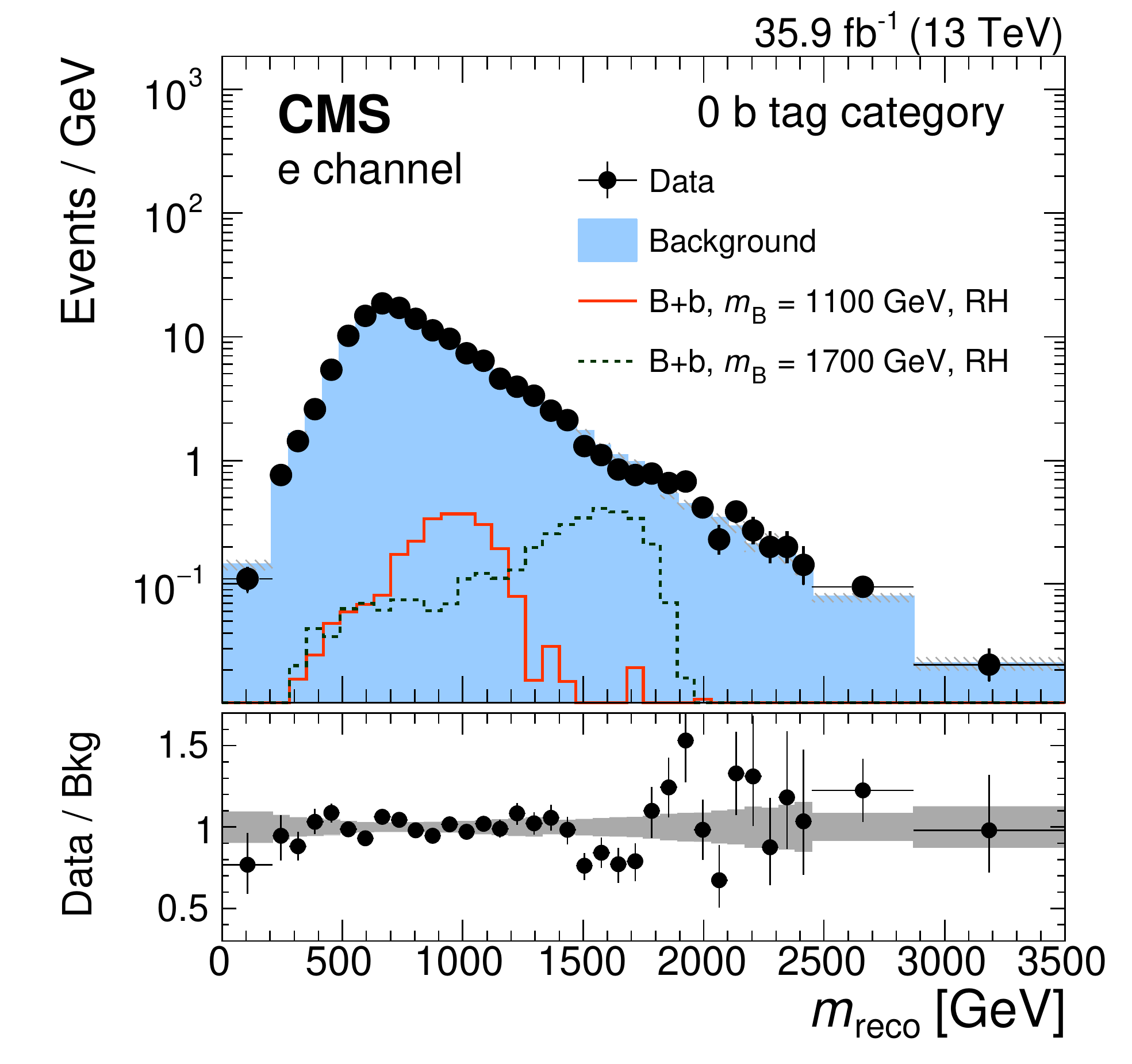}
 \caption{Distributions of \mreco\ measured in the signal region for
 events with a jet in the forward direction with $\abs{\eta}>2.4$ in the electron channel.
 Shown are the sensitive categories:
 \cPqt\ tag(upper left),
 \PW\ tag(upper right),
 ${\geq}2$ \cPqb\ tag (middle left),
 1 \cPqb\ tag (middle right) and 0 \cPqb\ tag (lower).
 The background prediction is obtained from control regions as
 detailed in the main text. The distributions from two example signal samples
 for the \PQB{}+\cPqb\ production mode with right-handed VLQ couplings
 with a cross section of 1\pb and a relative VLQ width of 1\% are shown for illustration.
 \label{fig:signalregion_ele}}
\end{figure*}

The measured distributions of \mreco for the signal
categories are shown in Figs.~\ref{fig:signalregion_mu}
and \ref{fig:signalregion_ele} for the muon and electron channels,
together with the background predictions obtained from the control regions.
The signal \mreco\ distributions for a vector-like \PQB quark with right-handed
couplings produced in association with a \cPqb\ quark are shown for illustration,
for two different VLQ masses with an assumed production cross section of 1\pb and
a relative VLQ width of 1\%. No significant deviation from the background expectation is observed in any of the categories.

Exclusion limits on the product of the VLQ production cross section and branching fraction
are calculated at 95\% confidence level (\CL) for VLQ masses
between 700 and 2000\GeV by using a Bayesian
statistical method~\cite{theta, bayesbook}.
Pseudo-experiments are performed to extract expected upper
limits under the background-only hypothesis.
For the signal cross section parameter an uniform prior distribution, and
for the nuisance parameters log-normal prior distributions are used.
The nuisance parameters are randomly varied within
their ranges of validity to estimate the 68 and 95\% \CL expected limits.
Correlations between the systematic uncertainties across all
channels are taken into account through a common nuisance parameter.
The statistical uncertainties of the background predictions are treated
as an additional Poisson nuisance parameter in each bin of the
\mreco\ distribution.

\begin{figure*}
 \centering
 \includegraphics[width=0.45\textwidth]{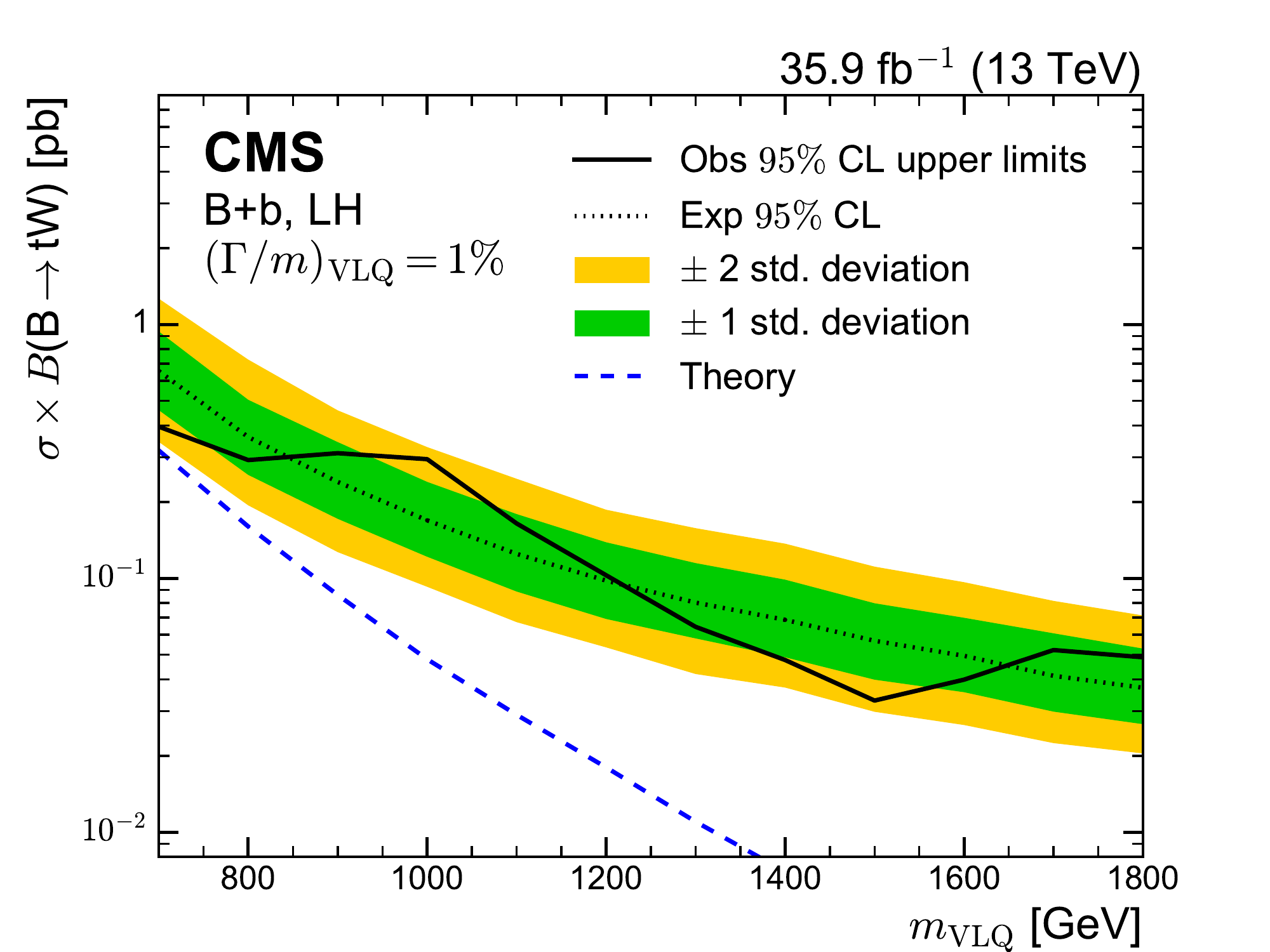}
 \includegraphics[width=0.45\textwidth]{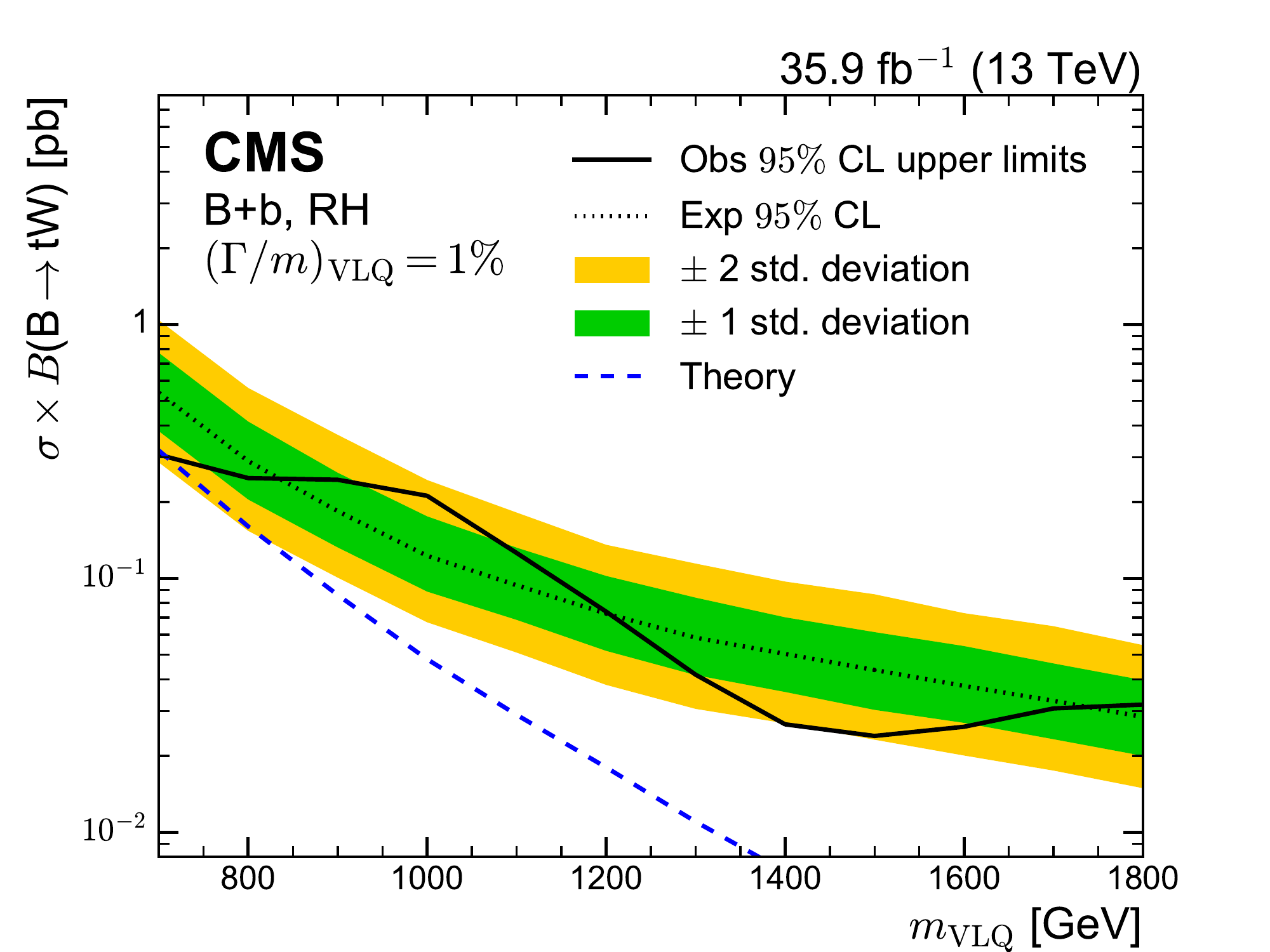}
 \includegraphics[width=0.45\textwidth]{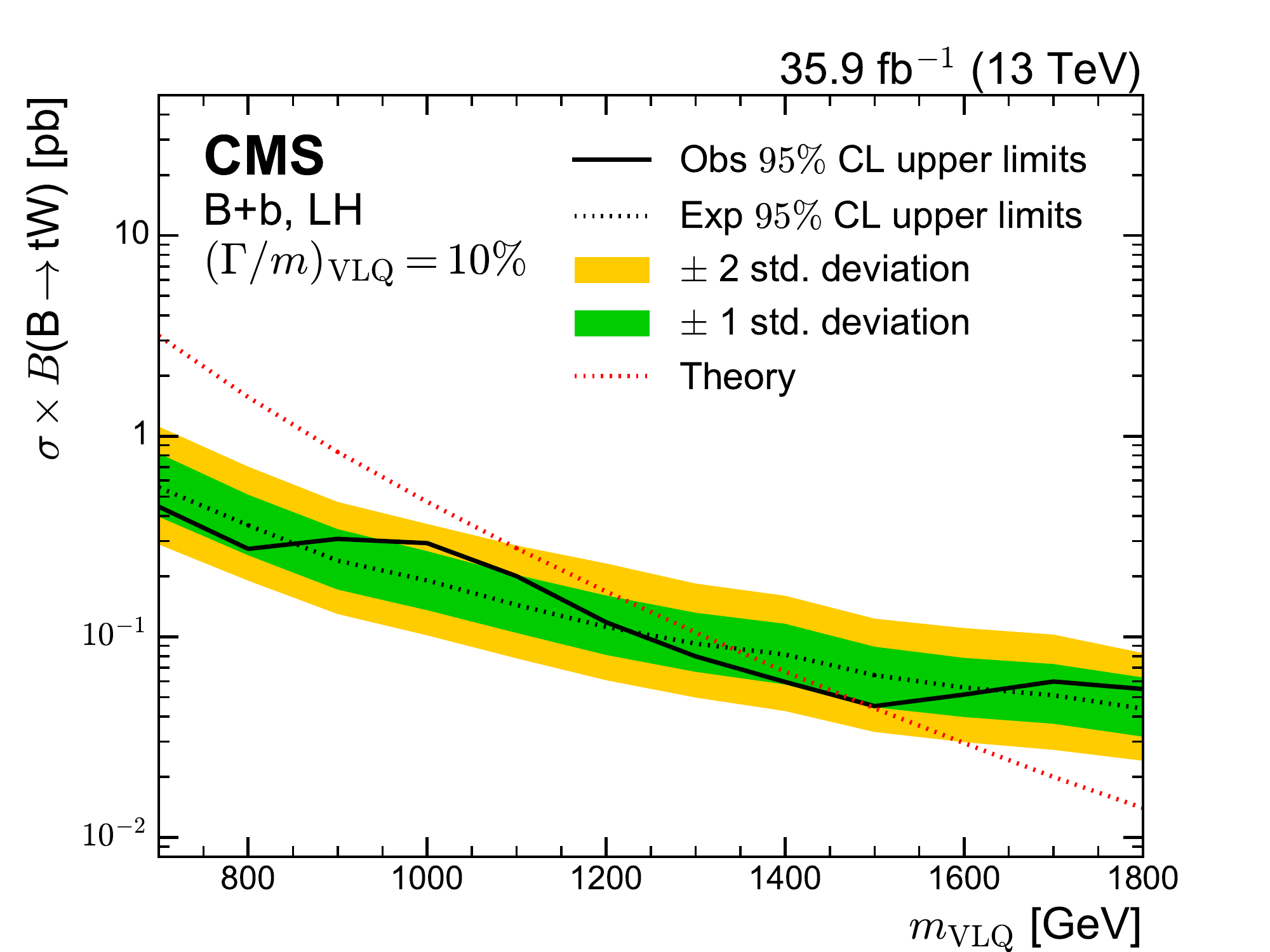}
 \includegraphics[width=0.45\textwidth]{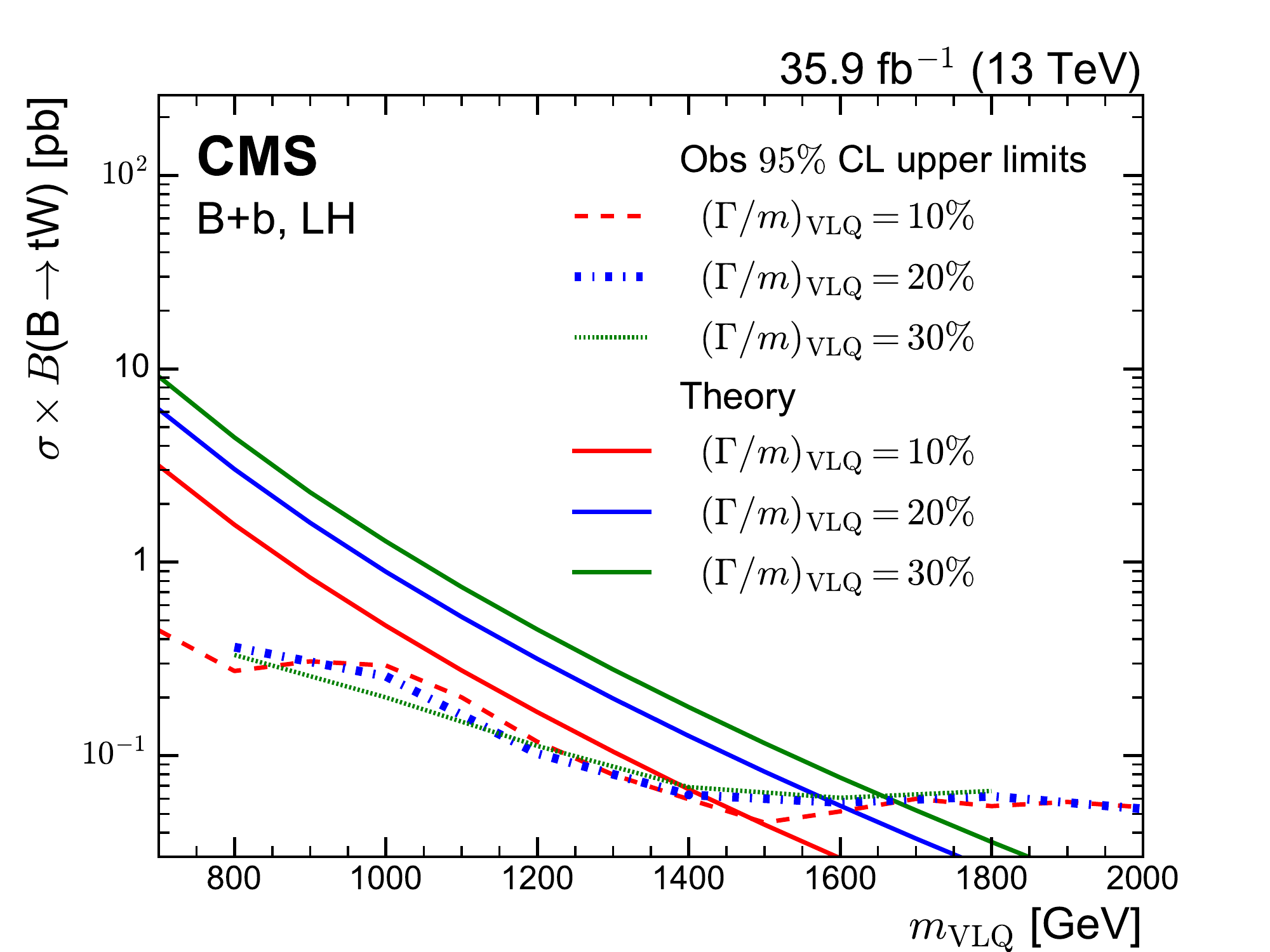}
 \caption{Upper limits at 95\% \CL on the product of the VLQ production cross section and
 branching fraction for the \PQB{}+\cPqb\ production mode for a relative VLQ width of 1\% and
 left- and right-handed VLQ couplings (upper left and right), for 10\% relative VLQ width
 and left-handed VLQ couplings (lower left), and
 a comparison of the observed exclusion limits for relative VLQ widths of 10, 20, and 30\%
 for left-handed couplings (lower right).
 The dashed lines show the theoretical predictions.
 \label{fig:limits_Bb} }
\end{figure*}

\begin{figure*}
 \centering
 \includegraphics[width=0.45\textwidth]{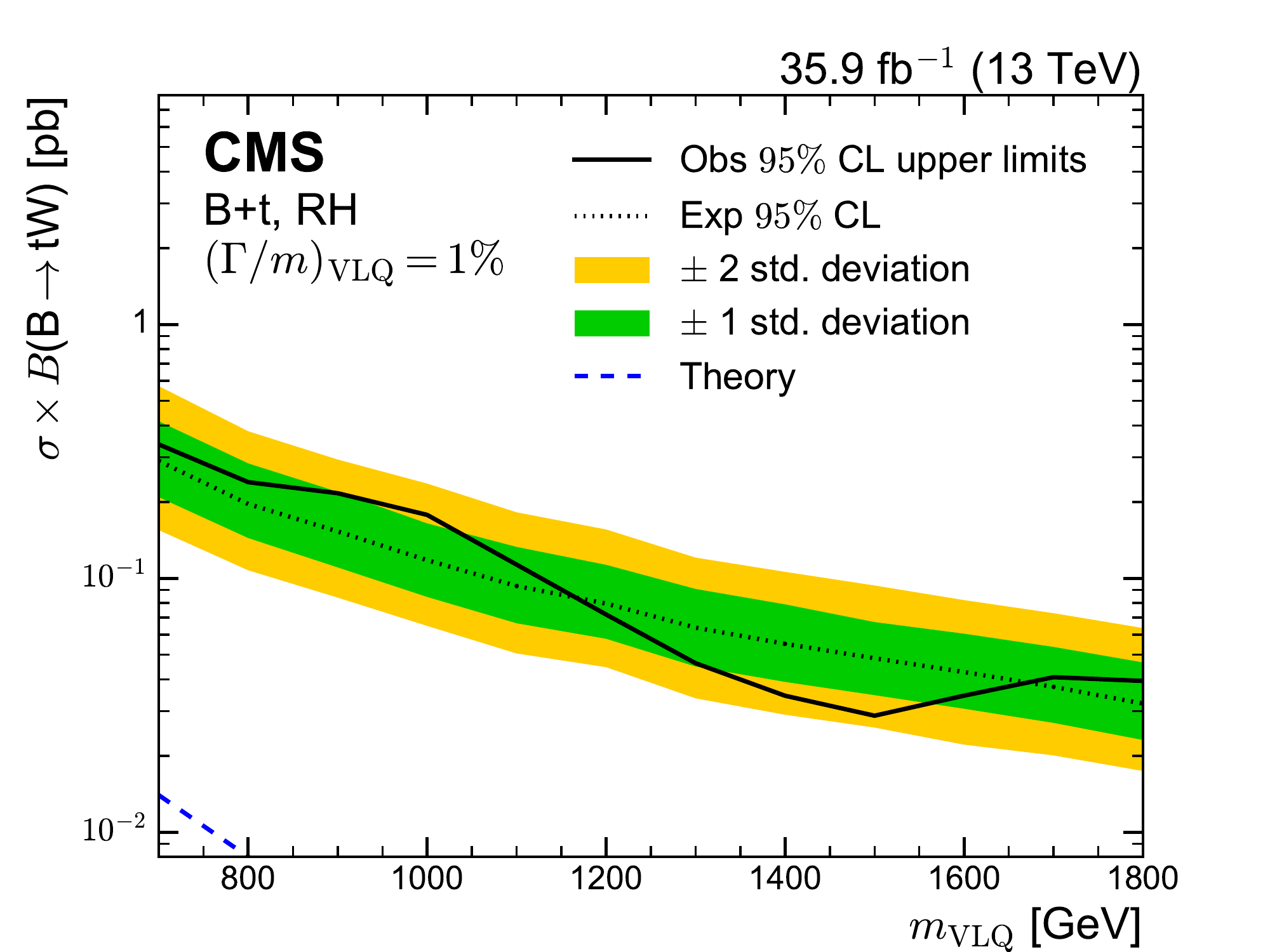}
 \includegraphics[width=0.45\textwidth]{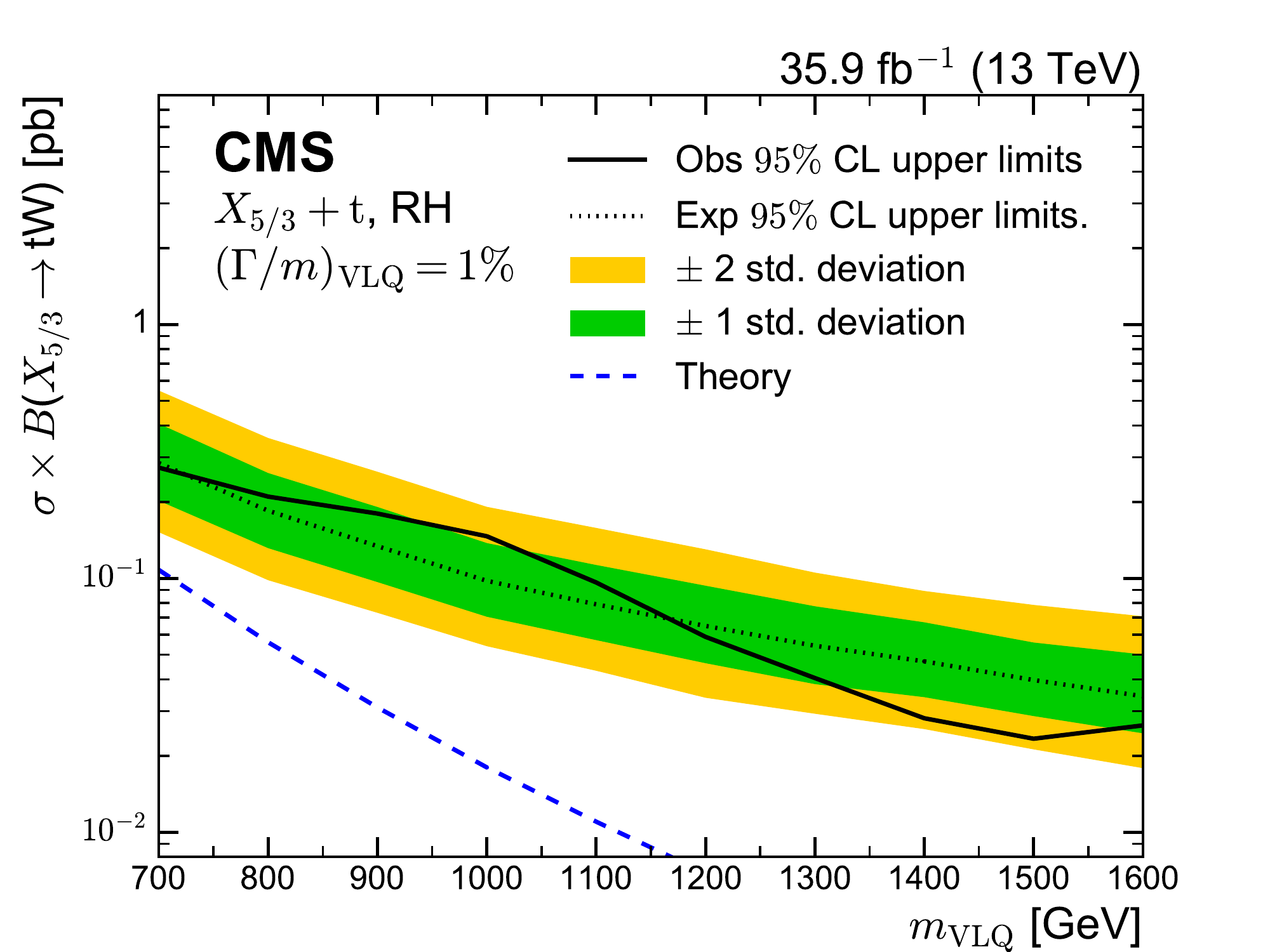}
 \includegraphics[width=0.45\textwidth]{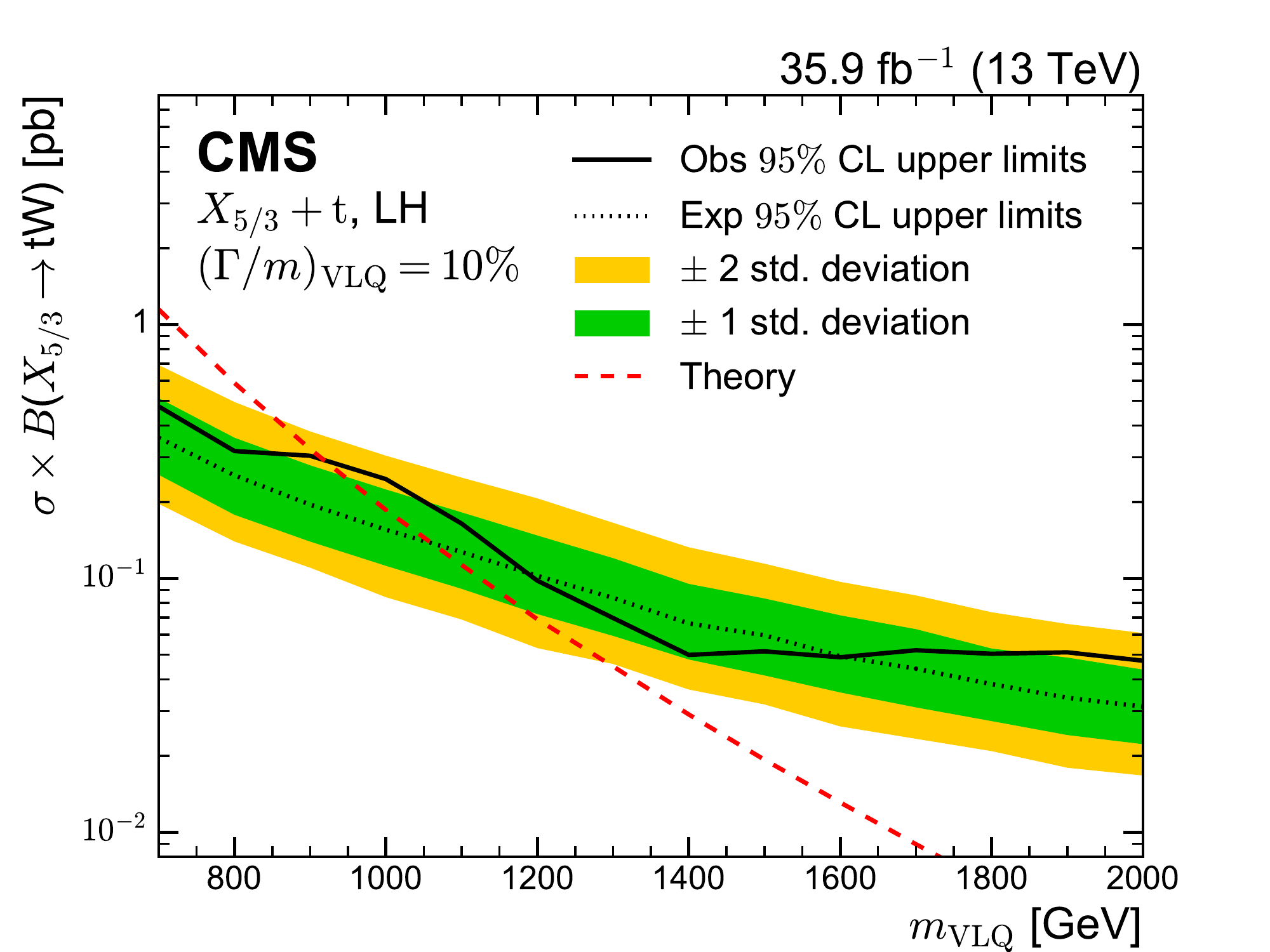}
 \includegraphics[width=0.45\textwidth]{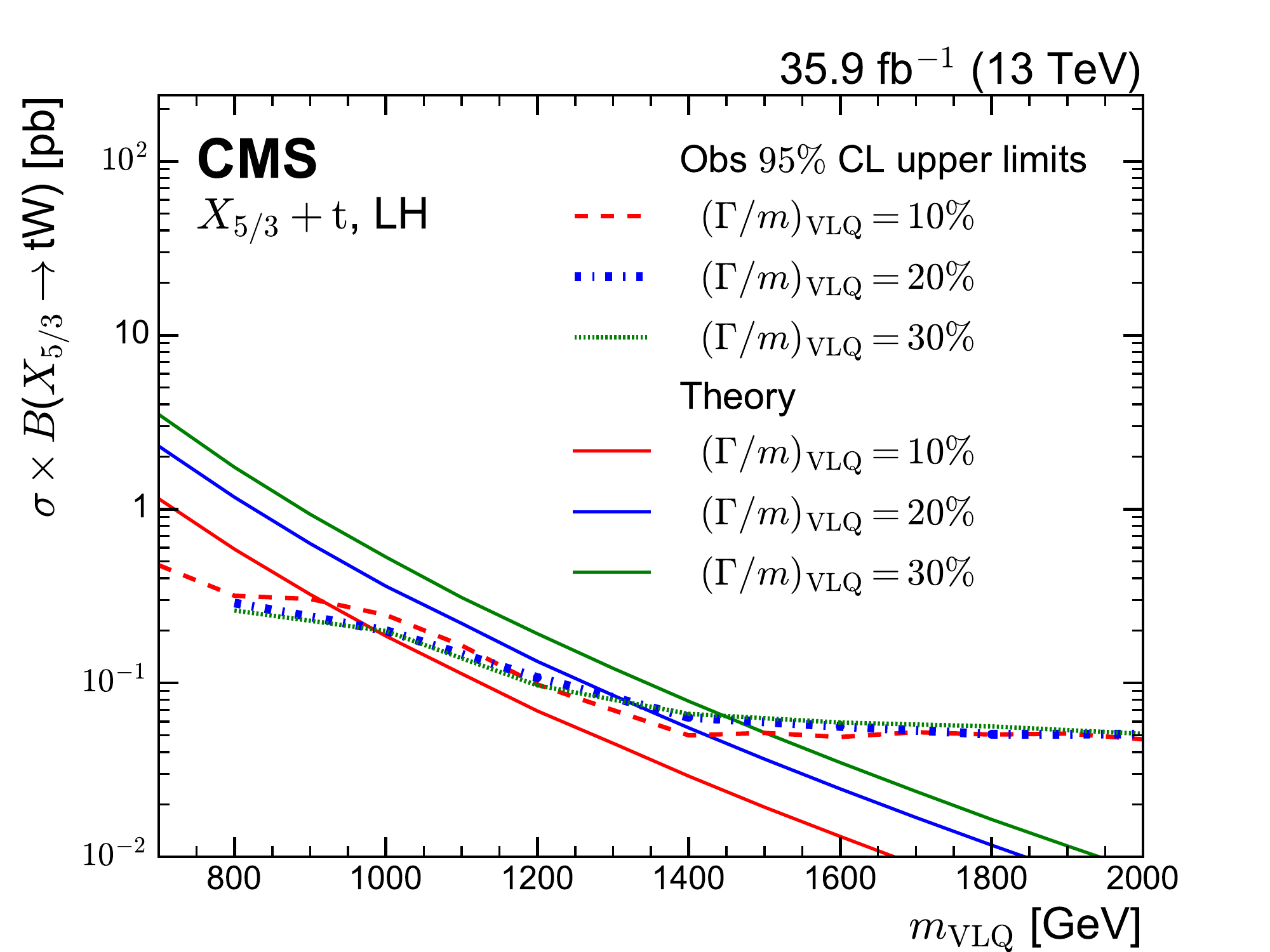}
 \caption{Upper limits at 95\% \CL on the product of the VLQ production
 cross section and branching fraction for the \PQB{}+\cPqt\ and \X+\cPqt\ production modes
 for right-handed VLQ couplings assuming a relative VLQ width of 1\% (upper left and right),
 for the \X+\cPqt\ production mode with left-handed VLQ couplings and a 10\% relative width (lower left) and a comparison of the observed exclusion
 limits for left-handed couplings for relative widths of 10, 20, and 30\% (lower right).
 The dashed lines show the theoretical predictions.
 \label{fig:limits_combined} }
\end{figure*}

Figure~\ref{fig:limits_Bb} shows the 95\% \CL upper limits on
the product of the cross section and branching fraction for the \PQB{}+\cPqb\ 
production mode for left- and right-handed VLQ couplings and a relative
VLQ width of 1\% (upper left and upper right), for the left-handed VLQ
couplings and a relative VLQ width of 10\% (lower left), as well as a
comparison of the observed exclusion limits for relative VLQ widths
between 10 and 30\% (lower right).
In Fig.~\ref{fig:limits_combined}, the 95\% \CL upper limits on the product of the cross section and branching fraction for the production
modes \PQB{}+\cPqt\ (upper left) and \X+\cPqt\ (upper right) and right-handed VLQ couplings are shown.
The figure also shows the \X+\cPqt\ exclusion limits for left-handed VLQ couplings with a 10\% relative VLQ width (lower left) and
a comparison of the observed exclusion limits for VLQ widths between 10 and 30\% for left-handed couplings (lower right).
The predicted cross sections for variations of the relative VLQ
mass width (dashed lines) are taken from Refs.~\cite{Carvalho:2018jkq,Matsedonskyi:2014mna,Campbell:2004ch}.
For a set of VLQ masses the expected and observed 95\% \CL upper limits for the \PQB{}+\cPqb\ and the \X+\cPqt\ production modes
are also given in Table~\ref{tab:mass_limits} for VLQs with widths of 1\% and 10\% and left-handed couplings, as well
as for widths of 1\% and right-handed couplings.
The exclusion limits for the \PQB{}+\cPqt\ production mode are similar to those for the \X+\cPqt\ production mode.

The obtained exclusion limits range from 0.3 to 0.03\pb for VLQ masses between 700 and 2000\GeV. For VLQs with a relative width of 1\% and purely left-handed
couplings an increase of about 25\% of the 95\% \CL upper limits is observed because of the reduced signal acceptance, in comparison to the right-handed couplings.
The expected limits for VLQ with relative widths of 10--30\% and left-handed couplings only show small differences. Although the predicted cross sections for the SM backgrounds are considerably larger at 13\TeV, similar exclusion limits on the product of cross section and branching fraction are achieved compared to the results obtained at 8\TeV in the more restricted mass range considered in Ref.~\cite{Aad:2015voa}. However, because of the increase of the VLQ signal cross section at 13\TeV, with this analysis, the existence of VLQ \PQB (\X) quarks with left-handed couplings and a relative width of 10, 20, and 30\% can be excluded for masses below 1490, 1590, and 1660\GeV (920, 1300, and 1450\GeV) respectively. The results represent the most stringent exclusion limits for singly produced VLQ in this channel.

\begin{table*}
  \centering
  \topcaption{Observed (expected) upper limits at 95\% \CL on the product of the cross section and branching fraction for the \PQB{}+\cPqb\ and \X+\cPqt\ production modes, for a set of VLQ masses, for VLQs widths of 1\% and 10\%, and for left-handed and right-handed couplings. The exclusion limits for the \PQB{}+\cPqt\ production mode (not shown) are very similar to those for the \X+\cPqt\ mode.}
  \cmsTable{
  \begin{tabular}{c c c c c c c}
     &  \multicolumn{3}{c}{\PQB{}+\cPqb} & \multicolumn{3}{c}{\X+\cPqt} \\
      \mVLQ [TeV] & 1\% LH& 10\% LH& 1\% RH& 1\% LH& 10\% LH& 1\% RH\\ \hline
      0.8    & 0.29 (0.36) & 0.27 (0.36) & 0.25 (0.29) & 0.31 (0.27) & 0.32 (0.25) & 0.21 (0.18) \\
      1      & 0.29 (0.17) & 0.29 (0.19) & 0.21 (0.12) & 0.25 (0.15) & 0.25 (0.16) & 0.15 (0.10) \\
      1.2    & 0.10 (0.10) & 0.11 (0.11) & 0.07 (0.07) & 0.10 (0.09) & 0.10 (0.10) & 0.06 (0.06) \\
      1.4    & 0.07 (0.07) & 0.06 (0.08) & 0.03 (0.05) & 0.05 (0.06) & 0.05 (0.07) & 0.03 (0.05) \\
      1.6    & 0.05 (0.05) & 0.05 (0.06) & 0.03 (0.04) & 0.04 (0.04) & 0.05 (0.05) & 0.03 (0.03) \\
      1.8    & 0.04 (0.04) & 0.05 (0.04) & 0.03 (0.03) & \NA         & 0.05 (0.04) & \NA         \\
  \end{tabular}
  }
  \label{tab:mass_limits}
\end{table*}

\section{Summary}
\label{sec:summary}
A search for singly produced vector-like quarks decaying into a top quark and a $\PW$
boson has been performed using the 2016 data set recorded by the CMS experiment
at the CERN LHC. The selection is optimised for high vector-like quark masses,
with a single muon or electron, significant missing transverse momentum,
and two jets with high $\pt$ in the final state.
Vector-like quarks in the single production mode can be produced in association with a $\cPqt$ or a $\cPqb$ quark
and a forward jet. The latter feature is used to obtain the background prediction
in the signal regions from data. The mass of the vector-like quark is reconstructed from the hadronic jets,
the missing transverse momentum, and the lepton in the event.
Different decay possibilities of the $\cPqt$ and $\PW$ are considered.
The reach of the search is enhanced by $\cPqt$, $\PW$, and $\cPqb$ tagging methods.
No significant deviation from the standard model prediction is observed.
Upper exclusion limits at 95\% confidence level on the product of the production cross section and
branching fraction range from around 0.3--0.03\pb for vector-like quark masses between 700 and 2000\GeV.
Depending on the vector-like quark type, coupling, and decay width to \cPqt\PW, mass exclusion limits up to 1660\GeV are obtained.
These represent the most stringent exclusion limits for the single production of vector-like quarks in this channel.

\begin{acknowledgments}

We congratulate our colleagues in the CERN accelerator departments for the excellent performance of the LHC and thank the technical and administrative staffs at CERN and at other CMS institutes for their contributions to the success of the CMS effort. In addition, we gratefully acknowledge the computing centres and personnel of the Worldwide LHC Computing Grid for delivering so effectively the computing infrastructure essential to our analyses. Finally, we acknowledge the enduring support for the construction and operation of the LHC and the CMS detector provided by the following funding agencies: BMWFW and FWF (Austria); FNRS and FWO (Belgium); CNPq, CAPES, FAPERJ, FAPERGS, and FAPESP (Brazil); MES (Bulgaria); CERN; CAS, MoST, and NSFC (China); COLCIENCIAS (Colombia); MSES and CSF (Croatia); RPF (Cyprus); SENESCYT (Ecuador); MoER, ERC IUT, and ERDF (Estonia); Academy of Finland, MEC, and HIP (Finland); CEA and CNRS/IN2P3 (France); BMBF, DFG, and HGF (Germany); GSRT (Greece); NKFIA (Hungary); DAE and DST (India); IPM (Iran); SFI (Ireland); INFN (Italy); MSIP and NRF (Republic of Korea); LAS (Lithuania); MOE and UM (Malaysia); BUAP, CINVESTAV, CONACYT, LNS, SEP, and UASLP-FAI (Mexico); MOS (Montenegro); MBIE (New Zealand); PAEC (Pakistan); MSHE and NSC (Poland); FCT (Portugal); JINR (Dubna); MON, RosAtom, RAS, RFBR, and NRC KI (Russia); MESTD (Serbia); SEIDI, CPAN, PCTI, and FEDER (Spain); MOSTR (Sri Lanka); Swiss Funding Agencies (Switzerland); MST (Taipei); ThEPCenter, IPST, STAR, and NSTDA (Thailand); TUBITAK and TAEK (Turkey); NASU and SFFR (Ukraine); STFC (United Kingdom); DOE and NSF (USA).

\hyphenation{Rachada-pisek} Individuals have received support from the Marie-Curie programme and the European Research Council and Horizon 2020 Grant, contract No. 675440 (European Union); the Leventis Foundation; the A. P. Sloan Foundation; the Alexander von Humboldt Foundation; the Belgian Federal Science Policy Office; the Fonds pour la Formation \`a la Recherche dans l'Industrie et dans l'Agriculture (FRIA-Belgium); the Agentschap voor Innovatie door Wetenschap en Technologie (IWT-Belgium); the F.R.S.-FNRS and FWO (Belgium) under the ``Excellence of Science - EOS" - be.h project n. 30820817; the Ministry of Education, Youth and Sports (MEYS) of the Czech Republic; the Lend\"ulet (``Momentum") Programme and the J\'anos Bolyai Research Scholarship of the Hungarian Academy of Sciences, the New National Excellence Program \'UNKP, the NKFIA research grants 123842, 123959, 124845, 124850 and 125105 (Hungary); the Council of Science and Industrial Research, India; the HOMING PLUS programme of the Foundation for Polish Science, cofinanced from European Union, Regional Development Fund, the Mobility Plus programme of the Ministry of Science and Higher Education, the National Science Center (Poland), contracts Harmonia 2014/14/M/ST2/00428, Opus 2014/13/B/ST2/02543, 2014/15/B/ST2/03998, and 2015/19/B/ST2/02861, Sonata-bis 2012/07/E/ST2/01406; the National Priorities Research Program by Qatar National Research Fund; the Programa Estatal de Fomento de la Investigaci{\'o}n Cient{\'i}fica y T{\'e}cnica de Excelencia Mar\'{\i}a de Maeztu, grant MDM-2015-0509 and the Programa Severo Ochoa del Principado de Asturias; the Thalis and Aristeia programmes cofinanced by EU-ESF and the Greek NSRF; the Rachadapisek Sompot Fund for Postdoctoral Fellowship, Chulalongkorn University and the Chulalongkorn Academic into Its 2nd Century Project Advancement Project (Thailand); the Welch Foundation, contract C-1845; and the Weston Havens Foundation (USA).

\end{acknowledgments}

\bibliography{auto_generated}

\providecommand{\href}[2]{#2}\begingroup\raggedright\begin{thebibliography}{10}%
\makeatletter
\providecommand{\hrefCMSnoop }[0]{\@secondoftwo}%
\makeatother
\providecommand{\doi}{\texttt{doi:}\begingroup \urlstyle{tt}\Url}

\bibitem{Aad:2012tfa}
\hrefCMSnoop {}{{ATLAS Collaboration}, ``Observation of a new particle in the
  search for the standard model {Higgs} boson with the {ATLAS} detector at the
  {LHC}'',} \textit{ Phys. Lett. B} \textbf{ 716} (2012) 1,
  \href{http://dx.doi.org/10.1016/j.physletb.2012.08.020}{\doi{10.1016/j.physletb.2012.08.020}},
\href{http://www.arXiv.org/abs/1207.7214}{\texttt{arXiv:1207.7214}}.
%%CITATION = ARXIV:1207.7214;%%.

\bibitem{Chatrchyan:2012xdj}
\hrefCMSnoop {}{{CMS Collaboration}, ``Observation of a new boson at a mass of
  125 {GeV} with the {CMS} experiment at the {LHC}'',} \textit{ Phys. Lett. B}
  \textbf{ 716} (2012) 30,
  \href{http://dx.doi.org/10.1016/j.physletb.2012.08.021}{\doi{10.1016/j.physletb.2012.08.021}},
\href{http://www.arXiv.org/abs/1207.7235}{\texttt{arXiv:1207.7235}}.
%%CITATION = ARXIV:1207.7235;%%.

\bibitem{ArkaniHamed:2002qy}
\hrefCMSnoop {}{N.~Arkani-Hamed, A.~G. Cohen, E.~Katz, and A.~E. Nelson, ``The
  littlest {Higgs}'',} \textit{ JHEP} \textbf{ 07} (2002) 034,
  \href{http://dx.doi.org/10.1088/1126-6708/2002/07/034}{\doi{10.1088/1126-6708/2002/07/034}},
\href{http://www.arXiv.org/abs/hep-ph/0206021}{\texttt{arXiv:hep-ph/0206021}}.
%%CITATION = HEP-PH/0206021;%%.

\bibitem{Schmaltz:2002wx}
\hrefCMSnoop {}{M.~Schmaltz, ``Physics beyond the standard model (theory):
  Introducing the little {Higgs}'',} \textit{ Nucl. Phys. Proc. Suppl.}
  \textbf{ 117} (2003) 40,
  \href{http://dx.doi.org/10.1016/S0920-5632(03)01409-9}{\doi{10.1016/S0920-5632(03)01409-9}},
\href{http://www.arXiv.org/abs/hep-ph/0210415}{\texttt{arXiv:hep-ph/0210415}}.
%%CITATION = HEP-PH/0210415;%%.

\bibitem{Schmaltz:2005ky}
\hrefCMSnoop {}{M.~Schmaltz and D.~Tucker-Smith, ``Little {Higgs} review'',}
  \textit{ Ann. Rev. Nucl. Part. Sci.} \textbf{ 55} (2005) 229,
  \href{http://dx.doi.org/10.1146/annurev.nucl.55.090704.151502}{\doi{10.1146/annurev.nucl.55.090704.151502}},
\href{http://www.arXiv.org/abs/hep-ph/0502182}{\texttt{arXiv:hep-ph/0502182}}.
%%CITATION = HEP-PH/0502182;%%.

\bibitem{Marzocca:2012zn}
\hrefCMSnoop {}{D.~Marzocca, M.~Serone, and J.~Shu, ``General composite {Higgs}
  models'',} \textit{ JHEP} \textbf{ 08} (2012) 013,
  \href{http://dx.doi.org/10.1007/JHEP08(2012)013}{\doi{10.1007/JHEP08(2012)013}},
\href{http://www.arXiv.org/abs/1205.0770}{\texttt{arXiv:1205.0770}}.
%%CITATION = ARXIV:1205.0770;%%.

\bibitem{Djouadi:2012ae}
\hrefCMSnoop {}{A.~Djouadi and A.~Lenz, ``Sealing the fate of a fourth
  generation of fermions'',} \textit{ Phys. Lett. B} \textbf{ 715} (2012) 310,
  \href{http://dx.doi.org/10.1016/j.physletb.2012.07.060}{\doi{10.1016/j.physletb.2012.07.060}},
\href{http://www.arXiv.org/abs/1204.1252}{\texttt{arXiv:1204.1252}}.
%%CITATION = ARXIV:1204.1252;%%.

\bibitem{VLQHandbook:2013}
\hrefCMSnoop {}{J.~A. Aguilar-Saavedra, R.~Benbrik, S.~Heinemeyer, and
  M.~P\'erez-Victoria, ``Handbook of vectorlike quarks: Mixing and single
  production'',} \textit{ Phys. Rev. D} \textbf{ 88} (2013) 094010,
  \href{http://dx.doi.org/10.1103/PhysRevD.88.094010}{\doi{10.1103/PhysRevD.88.094010}},
\href{http://www.arXiv.org/abs/1306.0572}{\texttt{arXiv:1306.0572}}.
%%CITATION = ARXIV:1306.0572;%%.

\bibitem{Khachatryan:2016vau}
\hrefCMSnoop {}{{ATLAS and CMS Collaborations}, ``Measurements of the {Higgs}
  boson production and decay rates and constraints on its couplings from a
  combined {ATLAS} and {CMS} analysis of the {LHC} {$\Pp\Pp$} collision data at
  {$ \sqrt{s}=7 $} and 8{ TeV}'',} \textit{ JHEP} \textbf{ 08} (2016) 045,
  \href{http://dx.doi.org/10.1007/JHEP08(2016)045}{\doi{10.1007/JHEP08(2016)045}},
\href{http://www.arXiv.org/abs/1606.02266}{\texttt{arXiv:1606.02266}}.
%%CITATION = ARXIV:1606.02266;%%.

\bibitem{Chatrchyan:2013uxa}
\hrefCMSnoop {}{{CMS Collaboration}, ``Inclusive search for a vector-like {T}
  quark with charge $\frac{2}{3}$ in pp collisions at {$\sqrt{s}$ = 8 TeV}'',}
  \textit{ Phys. Lett. B} \textbf{ 729} (2014) 149,
  \href{http://dx.doi.org/10.1016/j.physletb.2014.01.006}{\doi{10.1016/j.physletb.2014.01.006}},
\href{http://www.arXiv.org/abs/1311.7667}{\texttt{arXiv:1311.7667}}.
%%CITATION = ARXIV:1311.7667;%%.

\bibitem{Khachatryan:2015axa}
\hrefCMSnoop {}{{CMS Collaboration}, ``Search for vector-like {T} quarks
  decaying to top quarks and {Higgs} bosons in the all-hadronic channel using
  jet substructure'',} \textit{ JHEP} \textbf{ 06} (2015) 080,
  \href{http://dx.doi.org/10.1007/JHEP06(2015)080}{\doi{10.1007/JHEP06(2015)080}},
\href{http://www.arXiv.org/abs/1503.01952}{\texttt{arXiv:1503.01952}}.
%%CITATION = ARXIV:1503.01952;%%.

\bibitem{Khachatryan:2015oba}
\hrefCMSnoop {}{{CMS Collaboration}, ``Search for vector-like charge $2/3$ {T}
  quarks in proton-proton collisions at {$\sqrt{s} = 8 \TeV$}'',} \textit{
  Phys. Rev. D} \textbf{ 93} (2016) 012003,
  \href{http://dx.doi.org/10.1103/PhysRevD.93.012003}{\doi{10.1103/PhysRevD.93.012003}},
\href{http://www.arXiv.org/abs/1509.04177}{\texttt{arXiv:1509.04177}}.
%%CITATION = ARXIV:1509.04177;%%.

\bibitem{Khachatryan:2016vph}
\hrefCMSnoop {}{{CMS Collaboration}, ``Search for single production of a heavy
  vector-like {T} quark decaying to a {Higgs} boson and a top quark with a
  lepton and jets in the final state'',} \textit{ Phys. Lett. B} \textbf{ 771}
  (2017) 80,
  \href{http://dx.doi.org/10.1016/j.physletb.2017.05.019}{\doi{10.1016/j.physletb.2017.05.019}},
\href{http://www.arXiv.org/abs/1612.00999}{\texttt{arXiv:1612.00999}}.
%%CITATION = ARXIV:1612.00999;%%.

\bibitem{Sirunyan:2017tfc}
\hrefCMSnoop {}{{CMS Collaboration}, ``Search for single production of
  vector-like quarks decaying into a b quark and a {W} boson in proton-proton
  collisions at {$\sqrt s =$ 13 TeV}'',} \textit{ Phys. Lett. B} \textbf{ 772}
  (2017) 634,
  \href{http://dx.doi.org/10.1016/j.physletb.2017.07.022}{\doi{10.1016/j.physletb.2017.07.022}},
\href{http://www.arXiv.org/abs/1701.08328}{\texttt{arXiv:1701.08328}}.
%%CITATION = ARXIV:1701.08328;%%.

\bibitem{Sirunyan:2017jin}
\hrefCMSnoop {}{{CMS Collaboration}, ``Search for top quark partners with
  charge 5/3 in proton-proton collisions at {$ \sqrt{s}=13 $ TeV}'',} \textit{
  JHEP} \textbf{ 08} (2017) 073,
  \href{http://dx.doi.org/10.1007/JHEP08(2017)073}{\doi{10.1007/JHEP08(2017)073}},
\href{http://www.arXiv.org/abs/1705.10967}{\texttt{arXiv:1705.10967}}.
%%CITATION = ARXIV:1705.10967;%%.

\bibitem{Sirunyan:2017usq}
\hrefCMSnoop {}{{CMS Collaboration}, ``Search for pair production of
  vector-like {T} and {B} quarks in single-lepton final states using boosted
  jet substructure in proton-proton collisions at {$\sqrt{s}=13$ TeV}'',}
  \textit{ JHEP} \textbf{ 11} (2017) 085,
  \href{http://dx.doi.org/10.1007/JHEP11(2017)085}{\doi{10.1007/JHEP11(2017)085}},
\href{http://www.arXiv.org/abs/1706.03408}{\texttt{arXiv:1706.03408}}.
%%CITATION = ARXIV:1706.03408;%%.

\bibitem{Sirunyan:2017pks}
\hrefCMSnoop {}{{CMS Collaboration}, ``Search for pair production of
  vector-like quarks in the {bW$\overline{\mathrm{b}}$W} channel from
  proton-proton collisions at {$\sqrt{s} =$ 13 TeV}'',} \textit{ Phys. Lett. B}
  \textbf{ 779} (2018) 82,
  \href{http://dx.doi.org/10.1016/j.physletb.2018.01.077}{\doi{10.1016/j.physletb.2018.01.077}},
\href{http://www.arXiv.org/abs/1710.01539}{\texttt{arXiv:1710.01539}}.
%%CITATION = ARXIV:1710.01539;%%.

\bibitem{Sirunyan:2017ynj}
\hrefCMSnoop {}{{CMS Collaboration}, ``Search for single production of a
  vector-like {T} quark decaying to a {Z} boson and a top quark in
  proton-proton collisions at {$\sqrt{s} = 13\TeV$}'',} \textit{ Phys. Lett. B}
  \textbf{ 781} (2018) 574,
  \href{http://dx.doi.org/10.1016/j.physletb.2018.04.036}{\doi{10.1016/j.physletb.2018.04.036}},
\href{http://www.arXiv.org/abs/1708.01062}{\texttt{arXiv:1708.01062}}.
%%CITATION = ARXIV:1708.01062;%%.

\bibitem{Sirunyan:2018fjh}
\hrefCMSnoop {}{{CMS Collaboration}, ``Search for single production of
  vector-like quarks decaying to a b quark and a {Higgs} boson'',} \textit{
  JHEP} \textbf{ 06} (2018) 031,
  \href{http://dx.doi.org/10.1007/JHEP06(2018)031}{\doi{10.1007/JHEP06(2018)031}},
\href{http://www.arXiv.org/abs/1802.01486}{\texttt{arXiv:1802.01486}}.
%%CITATION = ARXIV:1802.01486;%%.

\bibitem{Aad:2011yn}
\hrefCMSnoop {}{{ATLAS Collaboration}, ``Search for heavy vector-like quarks
  coupling to light quarks in proton-proton collisions at {$\sqrt{s}=7$ TeV}
  with the {ATLAS} detector'',} \textit{ Phys. Lett. B} \textbf{ 712} (2012)
  22,
  \href{http://dx.doi.org/10.1016/j.physletb.2012.03.082}{\doi{10.1016/j.physletb.2012.03.082}},
\href{http://www.arXiv.org/abs/1112.5755}{\texttt{arXiv:1112.5755}}.
%%CITATION = ARXIV:1112.5755;%%.

\bibitem{Aad:2012bdq}
\hrefCMSnoop {}{{ATLAS Collaboration}, ``Search for pair production of a new
  quark that decays to a {Z} boson and a bottom quark with the {ATLAS}
  detector'',} \textit{ Phys. Rev. Lett.} \textbf{ 109} (2012) 071801,
  \href{http://dx.doi.org/10.1103/PhysRevLett.109.071801}{\doi{10.1103/PhysRevLett.109.071801}},
\href{http://www.arXiv.org/abs/1204.1265}{\texttt{arXiv:1204.1265}}.
%%CITATION = ARXIV:1204.1265;%%.

\bibitem{Aad:2014efa}
\hrefCMSnoop {}{{ATLAS Collaboration}, ``Search for pair and single production
  of new heavy quarks that decay to a {$Z$} boson and a third-generation quark
  in {$\Pp\Pp$} collisions at {$\sqrt{s}=8$ TeV} with the {ATLAS} detector'',}
  \textit{ JHEP} \textbf{ 11} (2014) 104,
  \href{http://dx.doi.org/10.1007/JHEP11(2014)104}{\doi{10.1007/JHEP11(2014)104}},
\href{http://www.arXiv.org/abs/1409.5500}{\texttt{arXiv:1409.5500}}.
%%CITATION = ARXIV:1409.5500;%%.

\bibitem{Aad:2015gdg}
\hrefCMSnoop {}{{ATLAS Collaboration}, ``Analysis of events with {$b$-jets} and
  a pair of leptons of the same charge in {$\Pp\Pp$} collisions at
  {$\sqrt{s}=8$ TeV} with the {ATLAS} detector'',} \textit{ JHEP} \textbf{ 10}
  (2015) 150,
  \href{http://dx.doi.org/10.1007/JHEP10(2015)150}{\doi{10.1007/JHEP10(2015)150}},
\href{http://www.arXiv.org/abs/1504.04605}{\texttt{arXiv:1504.04605}}.
%%CITATION = ARXIV:1504.04605;%%.

\bibitem{Aad:2015kqa}
\hrefCMSnoop {}{{ATLAS Collaboration}, ``Search for production of vector-like
  quark pairs and of four top quarks in the lepton-plus-jets final state in
  {$\Pp\Pp$} collisions at {$\sqrt{s}=8$ TeV} with the {ATLAS} detector'',}
  \textit{ JHEP} \textbf{ 08} (2015) 105,
  \href{http://dx.doi.org/10.1007/JHEP08(2015)105}{\doi{10.1007/JHEP08(2015)105}},
\href{http://www.arXiv.org/abs/1505.04306}{\texttt{arXiv:1505.04306}}.
%%CITATION = ARXIV:1505.04306;%%.

\bibitem{Aad:2015mba}
\hrefCMSnoop {}{{ATLAS Collaboration}, ``Search for vector-like {$B$} quarks in
  events with one isolated lepton, missing transverse momentum and jets at
  {$\sqrt{s}=$} 8 {TeV} with the {ATLAS} detector'',} \textit{ Phys. Rev. D}
  \textbf{ 91} (2015) 112011,
  \href{http://dx.doi.org/10.1103/PhysRevD.91.112011}{\doi{10.1103/PhysRevD.91.112011}},
\href{http://www.arXiv.org/abs/1503.05425}{\texttt{arXiv:1503.05425}}.
%%CITATION = ARXIV:1503.05425;%%.

\bibitem{Aaboud:2016lwz}
\hrefCMSnoop {}{{ATLAS Collaboration}, ``Search for top squarks in final states
  with one isolated lepton, jets, and missing transverse momentum in
  {$\sqrt{s}=13$ TeV $\Pp\Pp$} collisions with the {ATLAS} detector'',}
  \textit{ Phys. Rev. D} \textbf{ 94} (2016) 052009,
  \href{http://dx.doi.org/10.1103/PhysRevD.94.052009}{\doi{10.1103/PhysRevD.94.052009}},
\href{http://www.arXiv.org/abs/1606.03903}{\texttt{arXiv:1606.03903}}.
%%CITATION = ARXIV:1606.03903;%%.

\bibitem{Aad:2016qpo}
\hrefCMSnoop {}{{ATLAS Collaboration}, ``Search for single production of
  vector-like quarks decaying into {Wb} in {$\Pp\Pp$} collisions at {$\sqrt{s}
  = 8$ TeV} with the atlas detector'',} \textit{ Eur. Phys. J. C} \textbf{ 76}
  (2016) 442,
  \href{http://dx.doi.org/10.1140/epjc/s10052-016-4281-8}{\doi{10.1140/epjc/s10052-016-4281-8}},
\href{http://www.arXiv.org/abs/1602.05606}{\texttt{arXiv:1602.05606}}.
%%CITATION = ARXIV:1602.05606;%%.

\bibitem{Aad:2016shx}
\hrefCMSnoop {}{{ATLAS Collaboration}, ``Search for single production of a
  vector-like quark via a heavy gluon in the {$4b$} final state with the
  {ATLAS} detector in {$\Pp\Pp$} collisions at {$\sqrt{s} = 8$ TeV}'',}
  \textit{ Phys. Lett. B} \textbf{ 758} (2016) 249,
  \href{http://dx.doi.org/10.1016/j.physletb.2016.04.061}{\doi{10.1016/j.physletb.2016.04.061}},
\href{http://www.arXiv.org/abs/1602.06034}{\texttt{arXiv:1602.06034}}.
%%CITATION = ARXIV:1602.06034;%%.

\bibitem{Aaboud:2017qpr}
\hrefCMSnoop {}{{ATLAS Collaboration}, ``Search for pair production of
  vector-like top quarks in events with one lepton, jets, and missing
  transverse momentum in {$ \sqrt{s}=13 $ TeV $\Pp\Pp$} collisions with the
  {ATLAS} detector'',} \textit{ JHEP} \textbf{ 08} (2017) 052,
  \href{http://dx.doi.org/10.1007/JHEP08(2017)052}{\doi{10.1007/JHEP08(2017)052}},
\href{http://www.arXiv.org/abs/1705.10751}{\texttt{arXiv:1705.10751}}.
%%CITATION = ARXIV:1705.10751;%%.

\bibitem{Aaboud:2017zfn}
\hrefCMSnoop {}{{ATLAS Collaboration}, ``Search for pair production of heavy
  vector-like quarks decaying to high-{$p_{\mathrm{T}}$ W} bosons and b quarks
  in the lepton-plus-jets final state in {$\Pp\Pp$} collisions at {$
  \sqrt{s}=13 $ TeV} with the {ATLAS} detector'',} \textit{ JHEP} \textbf{ 10}
  (2017) 141,
  \href{http://dx.doi.org/10.1007/JHEP10(2017)141}{\doi{10.1007/JHEP10(2017)141}},
\href{http://www.arXiv.org/abs/1707.03347}{\texttt{arXiv:1707.03347}}.
%%CITATION = ARXIV:1707.03347;%%.

\bibitem{Aaboud:2018xuw}
\hrefCMSnoop {}{{ATLAS Collaboration}, ``Search for pair production of up-type
  vector-like quarks and for four-top-quark events in final states with
  multiple {$b$-jets} with the {ATLAS} detector'',} \textit{ JHEP} \textbf{ 07}
  (2018) 089,
  \href{http://dx.doi.org/10.1007/JHEP07(2018)089}{\doi{10.1007/JHEP07(2018)089}},
\href{http://www.arXiv.org/abs/1803.09678}{\texttt{arXiv:1803.09678}}.
%%CITATION = ARXIV:1803.09678;%%.

\bibitem{Aad:2015voa}
\hrefCMSnoop {}{{ATLAS Collaboration}, ``Search for the production of single
  vector-like and excited quarks in the {$\PW\cPqt$} final state in {$\Pp\Pp$}
  collisions at {$\sqrt{s}$ = 8 TeV} with the {ATLAS} detector'',} \textit{
  JHEP} \textbf{ 02} (2016) 110,
  \href{http://dx.doi.org/10.1007/JHEP02(2016)110}{\doi{10.1007/JHEP02(2016)110}},
\href{http://www.arXiv.org/abs/1510.02664}{\texttt{arXiv:1510.02664}}.
%%CITATION = ARXIV:1510.02664;%%.

\bibitem{Sirunyan:2017ulk}
\hrefCMSnoop {}{{CMS Collaboration}, ``Particle-flow reconstruction and global
  event description with the {CMS} detector'',} \textit{ JINST} \textbf{ 12}
  (2017) P10003,
  \href{http://dx.doi.org/10.1088/1748-0221/12/10/P10003}{\doi{10.1088/1748-0221/12/10/P10003}},
\href{http://www.arXiv.org/abs/1706.04965}{\texttt{arXiv:1706.04965}}.
%%CITATION = ARXIV:1706.04965;%%.

\bibitem{Khachatryan:2015hwa}
\hrefCMSnoop {}{{CMS Collaboration}, ``Performance of electron reconstruction
  and selection with the {CMS} detector in proton-proton collisions at
  $\sqrt{s} = 8$ {TeV}'',} \textit{ JINST} \textbf{ 10} (2015) P06005,
  \href{http://dx.doi.org/10.1088/1748-0221/10/06/P06005}{\doi{10.1088/1748-0221/10/06/P06005}},
\href{http://www.arXiv.org/abs/1502.02701}{\texttt{arXiv:1502.02701}}.
%%CITATION = ARXIV:1502.02701;%%.

\bibitem{Chatrchyan:2012xi}
\hrefCMSnoop {}{{CMS Collaboration}, ``Performance of {CMS} muon reconstruction
  in {$\Pp\Pp$} collision events at {$\sqrt{s} = 7$} {TeV}'',} \textit{ JINST}
  \textbf{ 7} (2012) P10002,
  \href{http://dx.doi.org/10.1088/1748-0221/7/10/P10002}{\doi{10.1088/1748-0221/7/10/P10002}},
\href{http://www.arXiv.org/abs/1206.4071}{\texttt{arXiv:1206.4071}}.
%%CITATION = ARXIV:1206.4071;%%.

\bibitem{Cacciari:2008gp}
\hrefCMSnoop {}{M.~Cacciari, G.~P. Salam, and G.~Soyez, ``The anti-{\kt} jet
  clustering algorithm'',} \textit{ JHEP} \textbf{ 04} (2008) 063,
  \href{http://dx.doi.org/10.1088/1126-6708/2008/04/063}{\doi{10.1088/1126-6708/2008/04/063}},
\href{http://www.arXiv.org/abs/0802.1189}{\texttt{arXiv:0802.1189}}.
%%CITATION = ARXIV:0802.1189;%%.

\bibitem{Cacciari:2011ma}
\hrefCMSnoop {}{M.~Cacciari, G.~P. Salam, and G.~Soyez, ``Fastjet user
  manual'',} \textit{ Eur. Phys. J. C} \textbf{ 72} (2012) 1896,
  \href{http://dx.doi.org/10.1140/epjc/s10052-012-1896-2}{\doi{10.1140/epjc/s10052-012-1896-2}},
\href{http://www.arXiv.org/abs/1111.6097}{\texttt{arXiv:1111.6097}}.
%%CITATION = ARXIV:1111.6097;%%.

\bibitem{Chatrchyan:2008zzk}
\hrefCMSnoop {}{{CMS Collaboration}, ``The {CMS} experiment at the {CERN}
  {LHC}'',} \textit{ JINST} \textbf{ 3} (2008) S08004,
\href{http://dx.doi.org/10.1088/1748-0221/3/08/S08004}{\doi{10.1088/1748-0221/3/08/S08004}}.
%%CITATION = JINST,3,S08004;%%.

\bibitem{Khachatryan:2016bia}
\hrefCMSnoop {}{{CMS Collaboration}, ``The {CMS} trigger system'',} \textit{
  JINST} \textbf{ 12} (2017) P01020,
  \href{http://dx.doi.org/10.1088/1748-0221/12/01/P01020}{\doi{10.1088/1748-0221/12/01/P01020}},
\href{http://www.arXiv.org/abs/1609.02366}{\texttt{arXiv:1609.02366}}.
%%CITATION = ARXIV:1609.02366;%%.

\bibitem{CMS-PAS-LUM-17-001}
\href {http://cds.cern.ch/record/2257069}{{CMS Collaboration}, ``{CMS}
  luminosity measurements for the 2016 data taking period'',} Technical Report
  CMS-PAS-LUM-17-001, 2017.

\bibitem{Nason:2004rx}
\hrefCMSnoop {}{P.~Nason, ``A new method for combining {NLO QCD} with shower
  {Monte Carlo} algorithms'',} \textit{ JHEP} \textbf{ 11} (2004) 040,
  \href{http://dx.doi.org/10.1088/1126-6708/2004/11/040}{\doi{10.1088/1126-6708/2004/11/040}},
\href{http://www.arXiv.org/abs/hep-ph/0409146}{\texttt{arXiv:hep-ph/0409146}}.
%%CITATION = HEP-PH/0409146;%%.

\bibitem{Frixione:2007vw}
\hrefCMSnoop {}{S.~Frixione, P.~Nason, and C.~Oleari, ``Matching {NLO QCD}
  computations with parton shower simulations: the {POWHEG} method'',} \textit{
  JHEP} \textbf{ 11} (2007) 070,
  \href{http://dx.doi.org/10.1088/1126-6708/2007/11/070}{\doi{10.1088/1126-6708/2007/11/070}},
\href{http://www.arXiv.org/abs/0709.2092}{\texttt{arXiv:0709.2092}}.
%%CITATION = ARXIV:0709.2092;%%.

\bibitem{Alioli:2010xd}
\hrefCMSnoop {}{S.~Alioli, P.~Nason, C.~Oleari, and E.~Re, ``A general
  framework for implementing nlo calculations in shower monte carlo programs:
  the {POWHEG BOX}'',} \textit{ JHEP} \textbf{ 06} (2010) 043,
  \href{http://dx.doi.org/10.1007/JHEP06(2010)043}{\doi{10.1007/JHEP06(2010)043}},
\href{http://www.arXiv.org/abs/1002.2581}{\texttt{arXiv:1002.2581}}.
%%CITATION = ARXIV:1002.2581;%%.

\bibitem{Alwall:2014hca}
J.~Alwall\hrefCMSnoop {}{ {et~al.}, ``The automated computation of tree-level
  and next-to-leading order differential cross sections, and their matching to
  parton shower simulations'',} \textit{ JHEP} \textbf{ 07} (2014) 079,
  \href{http://dx.doi.org/10.1007/JHEP07(2014)079}{\doi{10.1007/JHEP07(2014)079}},
\href{http://www.arXiv.org/abs/1405.0301}{\texttt{arXiv:1405.0301}}.
%%CITATION = ARXIV:1405.0301;%%.

\bibitem{Frederix:2012ps}
\hrefCMSnoop {}{R.~Frederix and S.~Frixione, ``Merging meets matching in
  {MC@NLO}'',} \textit{ JHEP} \textbf{ 12} (2012) 061,
  \href{http://dx.doi.org/10.1007/JHEP12(2012)061}{\doi{10.1007/JHEP12(2012)061}},
\href{http://www.arXiv.org/abs/1209.6215}{\texttt{arXiv:1209.6215}}.
%%CITATION = ARXIV:1209.6215;%%.

\bibitem{Alwall:2007fs}
\hrefCMSnoop {}{J.~Alwall {et~al.}, ``Comparative study of various algorithms
  for the merging of parton showers and matrix elements in hadronic
  collisions'',} \textit{ Eur. Phys. J. C} \textbf{ 53} (2008) 473,
  \href{http://dx.doi.org/10.1140/epjc/s10052-007-0490-5}{\doi{10.1140/epjc/s10052-007-0490-5}},
\href{http://www.arXiv.org/abs/0706.2569}{\texttt{arXiv:0706.2569}}.
%%CITATION = ARXIV:0706.2569;%%.

\bibitem{Sjostrand:2014zea}
T.~Sj{\"o}strand\hrefCMSnoop {}{ {et~al.}, ``An introduction to {PYTHIA
  8.2}'',} \textit{ Comput. Phys. Commun.} \textbf{ 191} (2015) 159,
  \href{http://dx.doi.org/10.1016/j.cpc.2015.01.024}{\doi{10.1016/j.cpc.2015.01.024}},
\href{http://www.arXiv.org/abs/1410.3012}{\texttt{arXiv:1410.3012}}.
%%CITATION = ARXIV:1410.3012;%%.

\bibitem{Ball:2014uwa}
\hrefCMSnoop {}{{NNPDF} Collaboration, ``Parton distributions for the {LHC} run
  {II}'',} \textit{ JHEP} \textbf{ 04} (2015) 040,
  \href{http://dx.doi.org/10.1007/JHEP04(2015)040}{\doi{10.1007/JHEP04(2015)040}},
\href{http://www.arXiv.org/abs/1410.8849}{\texttt{arXiv:1410.8849}}.
%%CITATION = ARXIV:1410.8849;%%.

\bibitem{Khachatryan:2015pea}
\hrefCMSnoop {}{{CMS Collaboration}, ``Event generator tunes obtained from
  underlying event and multiparton scattering measurements'',} \textit{ Eur.
  Phys. J. C} \textbf{ 76} (2016) 155,
  \href{http://dx.doi.org/10.1140/epjc/s10052-016-3988-x}{\doi{10.1140/epjc/s10052-016-3988-x}},
\href{http://www.arXiv.org/abs/1512.00815}{\texttt{arXiv:1512.00815}}.
%%CITATION = ARXIV:1512.00815;%%.

\bibitem{Skands2014}
\hrefCMSnoop {}{P.~Skands, S.~Carrazza, and J.~Rojo, ``Tuning {PYTHIA 8.1}: the
  {Monash} 2013 tune'',} \textit{ Eur. Phys. J. C} \textbf{ 74} (2014) 3024,
  \href{http://dx.doi.org/10.1140/epjc/s10052-014-3024-y}{\doi{10.1140/epjc/s10052-014-3024-y}},
\href{http://www.arXiv.org/abs/1404.5630}{\texttt{arXiv:1404.5630}}.
%%CITATION = ARXIV:1404.5630;%%.

\bibitem{CMS-PAS-TOP-16-021}
\href {https://cds.cern.ch/record/2235192}{{CMS Collaboration},
  ``Investigations of the impact of the parton shower tuning in {Pythia} 8 in
  the modelling of {$\mathrm{t\overline{t}}$} at {$\sqrt{s}=8$} and 13
  {TeV}'',} Technical Report CMS-PAS-TOP-16-021, CERN, Geneva, 2016.

\bibitem{Buchkremer:2013bha}
\hrefCMSnoop {}{M.~Buchkremer, G.~Cacciapaglia, A.~Deandrea, and L.~Panizzi,
  ``Model independent framework for searches of top partners'',} \textit{ Nucl.
  Phys. B} \textbf{ 876} (2013) 376,
  \href{http://dx.doi.org/10.1016/j.nuclphysb.2013.08.010}{\doi{10.1016/j.nuclphysb.2013.08.010}},
\href{http://www.arXiv.org/abs/1305.4172}{\texttt{arXiv:1305.4172}}.
%%CITATION = ARXIV:1305.4172;%%.

\bibitem{Fuks:2016ftf}
\hrefCMSnoop {}{B.~Fuks and H.-S. Shao, ``{QCD} next-to-leading-order
  predictions matched to parton showers for vector-like quark models'',}
  \textit{ Eur. Phys. J. C} \textbf{ 77} (2017) 135,
  \href{http://dx.doi.org/10.1140/epjc/s10052-017-4686-z}{\doi{10.1140/epjc/s10052-017-4686-z}},
\href{http://www.arXiv.org/abs/1610.04622}{\texttt{arXiv:1610.04622}}.
%%CITATION = ARXIV:1610.04622;%%.

\bibitem{Oliveira:2014kla}
\hrefCMSnoop {}{A.~Carvalho, ``Gravity particles from warped extra dimensions,
  predictions for {LHC}'',} (2014).
\href{http://www.arXiv.org/abs/1404.0102}{\texttt{arXiv:1404.0102}}.
%%CITATION = ARXIV:1404.0102;%%.

\bibitem{Campbell:2004ch}
\hrefCMSnoop {}{J.~M. Campbell, R.~K. Ellis, and F.~Tramontano, ``Single top
  production and decay at next-to-leading order'',} \textit{ Phys. Rev. D}
  \textbf{ 70} (2004) 094012,
  \href{http://dx.doi.org/10.1103/PhysRevD.70.094012}{\doi{10.1103/PhysRevD.70.094012}},
\href{http://www.arXiv.org/abs/hep-ph/0408158}{\texttt{arXiv:hep-ph/0408158}}.
%%CITATION = HEP-PH/0408158;%%.

\bibitem{Matsedonskyi:2014mna}
\hrefCMSnoop {}{O.~Matsedonskyi, G.~Panico, and A.~Wulzer, ``On the
  interpretation of top partners searches'',} \textit{ JHEP} \textbf{ 12}
  (2014) 097,
  \href{http://dx.doi.org/10.1007/JHEP12(2014)097}{\doi{10.1007/JHEP12(2014)097}},
\href{http://www.arXiv.org/abs/1409.0100}{\texttt{arXiv:1409.0100}}.
%%CITATION = ARXIV:1409.0100;%%.

\bibitem{Carvalho:2018jkq}
A.~Carvalho\hrefCMSnoop {}{ {et~al.}, ``Single production of vector-like quarks
  with large width at the {Large Hadron Collider}'',} \textit{ Phys. Rev. D}
  \textbf{ 98} (2018) 015029,
  \href{http://dx.doi.org/10.1103/PhysRevD.98.015029}{\doi{10.1103/PhysRevD.98.015029}},
\href{http://www.arXiv.org/abs/1805.06402}{\texttt{arXiv:1805.06402}}.
%%CITATION = ARXIV:1805.06402;%%.

\bibitem{Frixione:2007zp}
\hrefCMSnoop {}{S.~Frixione, E.~Laenen, P.~Motylinski, and B.~R. Webber,
  ``Angular correlations of lepton pairs from vector boson and top quark decays
  in {Monte Carlo} simulations'',} \textit{ JHEP} \textbf{ 04} (2007) 081,
  \href{http://dx.doi.org/10.1088/1126-6708/2007/04/081}{\doi{10.1088/1126-6708/2007/04/081}},
\href{http://www.arXiv.org/abs/hep-ph/0702198}{\texttt{arXiv:hep-ph/0702198}}.
%%CITATION = HEP-PH/0702198;%%.

\bibitem{Artoisenet:2012st}
\hrefCMSnoop {}{P.~Artoisenet, R.~Frederix, O.~Mattelaer, and R.~Rietkerk,
  ``Automatic spin-entangled decays of heavy resonances in {Monte Carlo}
  simulations'',} \textit{ JHEP} \textbf{ 03} (2013) 015,
  \href{http://dx.doi.org/10.1007/JHEP03(2013)015}{\doi{10.1007/JHEP03(2013)015}},
\href{http://www.arXiv.org/abs/1212.3460}{\texttt{arXiv:1212.3460}}.
%%CITATION = ARXIV:1212.3460;%%.

\bibitem{AGOSTINELLI2003250}
\hrefCMSnoop {}{{{\GEANTfour}} Collaboration, ``{\GEANTfour}--a simulation
  toolkit'',} \textit{ Nucl. Instrum. Meth. A} \textbf{ 506} (2003) 250,
\href{http://dx.doi.org/10.1016/S0168-9002(03)01368-8}{\doi{10.1016/S0168-9002(03)01368-8}}.
%%CITATION = NUIMA,A506,250;%%.

\bibitem{Khachatryan:2016kdb}
\hrefCMSnoop {}{{CMS Collaboration}, ``Jet energy scale and resolution in the
  {CMS} experiment in {$\Pp\Pp$} collisions at 8 {TeV}'',} \textit{ JINST}
  \textbf{ 12} (2017) P02014,
  \href{http://dx.doi.org/10.1088/1748-0221/12/02/P02014}{\doi{10.1088/1748-0221/12/02/P02014}},
\href{http://www.arXiv.org/abs/1607.03663}{\texttt{arXiv:1607.03663}}.
%%CITATION = ARXIV:1607.03663;%%.

\bibitem{Sirunyan:2018fpa}
\hrefCMSnoop {}{{CMS Collaboration}, ``Performance of the {CMS} muon detector
  and muon reconstruction with proton-proton collisions at {$\sqrt{s}=13$}
  {TeV}'',} \textit{ JINST} \textbf{ 13} (2018) P06015,
  \href{http://dx.doi.org/10.1088/1748-0221/13/06/P06015}{\doi{10.1088/1748-0221/13/06/P06015}},
\href{http://www.arXiv.org/abs/1804.04528}{\texttt{arXiv:1804.04528}}.
%%CITATION = ARXIV:1804.04528;%%.

\bibitem{CMS:2016tvk}
\href {https://cds.cern.ch/record/2126325}{{CMS Collaboration}, ``Top tagging
  with new approaches'',} Technical Report CMS-PAS-JME-15-002, CERN, Geneva,
  2016.

\bibitem{CMS-PAS-JME-16-003}
\href {https://cds.cern.ch/record/2256875}{{CMS Collaboration}, ``Jet
  algorithms performance in 13 {TeV} data'',} Technical Report
  CMS-PAS-JME-16-003, 2017.

\bibitem{Dasgupta:2013ihk}
\hrefCMSnoop {}{M.~Dasgupta, A.~Fregoso, S.~Marzani, and G.~P. Salam, ``Towards
  an understanding of jet substructure'',} \textit{ JHEP} \textbf{ 09} (2013)
  029,
  \href{http://dx.doi.org/10.1007/JHEP09(2013)029}{\doi{10.1007/JHEP09(2013)029}},
\href{http://www.arXiv.org/abs/1307.0007}{\texttt{arXiv:1307.0007}}.
%%CITATION = ARXIV:1307.0007;%%.

\bibitem{Larkoski:2014wba}
\hrefCMSnoop {}{A.~J. Larkoski, S.~Marzani, G.~Soyez, and J.~Thaler, ``{Soft
  Drop}'',} \textit{ JHEP} \textbf{ 05} (2014) 146,
  \href{http://dx.doi.org/10.1007/JHEP05(2014)146}{\doi{10.1007/JHEP05(2014)146}},
\href{http://www.arXiv.org/abs/1402.2657}{\texttt{arXiv:1402.2657}}.
%%CITATION = ARXIV:1402.2657;%%.

\bibitem{Sirunyan:2017ezt}
\hrefCMSnoop {}{{CMS Collaboration}, ``Identification of heavy-flavour jets
  with the {CMS} detector in pp collisions at {13 TeV}'',} \textit{ JINST}
  \textbf{ 13} (2018) P05011,
  \href{http://dx.doi.org/10.1088/1748-0221/13/05/P05011}{\doi{10.1088/1748-0221/13/05/P05011}},
\href{http://www.arXiv.org/abs/1712.07158}{\texttt{arXiv:1712.07158}}.
%%CITATION = ARXIV:1712.07158;%%.

\bibitem{Thaler:2010tr}
\hrefCMSnoop {}{J.~Thaler and K.~Van~Tilburg, ``Identifying boosted objects
  with {N-subjettiness}'',} \textit{ JHEP} \textbf{ 03} (2011) 015,
  \href{http://dx.doi.org/10.1007/JHEP03(2011)015}{\doi{10.1007/JHEP03(2011)015}},
\href{http://www.arXiv.org/abs/1011.2268}{\texttt{arXiv:1011.2268}}.
%%CITATION = ARXIV:1011.2268;%%.

\bibitem{Thaler:2011gf}
\hrefCMSnoop {}{J.~Thaler and K.~Van~Tilburg, ``Maximizing boosted top
  identification by minimizing {N-subjettiness}'',} \textit{ JHEP} \textbf{ 02}
  (2012) 093,
  \href{http://dx.doi.org/10.1007/JHEP02(2012)093}{\doi{10.1007/JHEP02(2012)093}},
\href{http://www.arXiv.org/abs/1108.2701}{\texttt{arXiv:1108.2701}}.
%%CITATION = ARXIV:1108.2701;%%.

\bibitem{Chatrchyan:2012nj}
\hrefCMSnoop {}{{CMS Collaboration}, ``Measurement of the inelastic
  proton-proton cross section at {$\sqrt{s}=7$ TeV}'',} \textit{ Phys. Lett. B}
  \textbf{ 722} (2013) 5,
  \href{http://dx.doi.org/10.1016/j.physletb.2013.03.024}{\doi{10.1016/j.physletb.2013.03.024}},
\href{http://www.arXiv.org/abs/1210.6718}{\texttt{arXiv:1210.6718}}.
%%CITATION = ARXIV:1210.6718;%%.

\bibitem{Butterworth:2015oua}
\hrefCMSnoop {}{J.~Butterworth {et~al.}, ``{PDF4LHC} recommendations for {LHC}
  run {II}'',} \textit{ J. Phys. G} \textbf{ 43} (2016) 023001,
  \href{http://dx.doi.org/10.1088/0954-3899/43/2/023001}{\doi{10.1088/0954-3899/43/2/023001}},
\href{http://www.arXiv.org/abs/1510.03865}{\texttt{arXiv:1510.03865}}.
%%CITATION = ARXIV:1510.03865;%%.

\bibitem{Bahr:2008pv}
M.~B{\"a}hr\hrefCMSnoop {}{ {et~al.}, ``{Herwig++} physics and manual'',}
  \textit{ Eur. Phys. J. C} \textbf{ 58} (2008) 639,
  \href{http://dx.doi.org/10.1140/epjc/s10052-008-0798-9}{\doi{10.1140/epjc/s10052-008-0798-9}},
\href{http://www.arXiv.org/abs/0803.0883}{\texttt{arXiv:0803.0883}}.
%%CITATION = ARXIV:0803.0883;%%.

\bibitem{theta}
\hrefCMSnoop {}{J.~Ott}, ``\textsc{Theta} --- {A} framework for template-based
  modeling and inference'' (2010),
  \url{http://www-ekp.physik.uni-karlsruhe.de/\~ott/theta/theta-auto}.

\bibitem{bayesbook}
A.~O'Hagan and J.~J. Forster, ``Kendall's advanced theory of statistics. {Vol.
  2B: Bayesian Inference}''.
\newblock Arnold, London, 2004.
\newblock ISBN~978-0470685693.

\end{thebibliography}\endgroup
\cleardoublepage \appendix\section{The CMS Collaboration \label{app:collab}}\begin{sloppypar}\hyphenpenalty=5000\widowpenalty=500\clubpenalty=5000\vskip\cmsinstskip
\textbf{Yerevan Physics Institute, Yerevan, Armenia}\\*[0pt]
A.M.~Sirunyan, A.~Tumasyan
\vskip\cmsinstskip
\textbf{Institut f\"{u}r Hochenergiephysik, Wien, Austria}\\*[0pt]
W.~Adam, F.~Ambrogi, E.~Asilar, T.~Bergauer, J.~Brandstetter, M.~Dragicevic, J.~Er\"{o}, A.~Escalante~Del~Valle, M.~Flechl, R.~Fr\"{u}hwirth\cmsAuthorMark{1}, V.M.~Ghete, J.~Hrubec, M.~Jeitler\cmsAuthorMark{1}, N.~Krammer, I.~Kr\"{a}tschmer, D.~Liko, T.~Madlener, I.~Mikulec, N.~Rad, H.~Rohringer, J.~Schieck\cmsAuthorMark{1}, R.~Sch\"{o}fbeck, M.~Spanring, D.~Spitzbart, A.~Taurok, W.~Waltenberger, J.~Wittmann, C.-E.~Wulz\cmsAuthorMark{1}, M.~Zarucki
% ...
\vskip\cmsinstskip
% ...
1:  Also at Vienna University of Technology, Vienna, Austria\\
% ...
\end{sloppypar}
\end{document}